\documentclass{article}

\usepackage{arxiv}

\usepackage{natbib}
\usepackage[utf8]{inputenc} 
\usepackage[T1]{fontenc}    
\usepackage{hyperref}       
\usepackage{url}            
\usepackage{booktabs}       
\usepackage{amsfonts}       
\usepackage{nicefrac}       
\usepackage{microtype}      
\usepackage{lipsum}
\usepackage{graphicx}
\graphicspath{ {./images/} }

\usepackage{array}
\usepackage{multirow}
\usepackage{lipsum} 
\usepackage{graphicx}
\usepackage{colortbl} 
\usepackage{xcolor}   
\usepackage{booktabs}
\usepackage{adjustbox}
\usepackage{tabularx}
\usepackage{caption}
\usepackage{soul}

\title{Capturing and Sharing Know-How through Visual Process Representations: A Human-Centred Approach to Teacher Workflows}

\author{
 Gloria Fernández-Nieto \\
  Faculty of Information Technology\\
  Monash University\\
  Melbourne, VIC, Australia \\
  \texttt{gloriamilena.fernandeznieto@monash.edu} \\
   \And
   Vanessa Echeverria \\
  Faculty of Information Technology\\
  Monash University\\
  Melbourne, VIC, Australia \\
  \texttt{vanessa.echeverria@monash.edu} \\
  \And
  Yuheng Li \\
  Faculty of Information Technology\\
  Monash University\\
  Melbourne, VIC, Australia \\
  \texttt{yuheng.li@monash.edu} \\
  \And
  Yi-Shan Tsai\\
  Faculty of Information Technology\\
  Monash University\\
  Melbourne, VIC, Australia \\
  \texttt{yuheng.li@monash.edu} \\
  \And
  Lele Sha\\
  Faculty of Information Technology\\
  Monash University\\
  Melbourne, VIC, Australia \\
  \texttt{lele.sha@monash.edu} \\
  \And
  Guanliang Chen\\
  Faculty of Information Technology\\
  Monash University\\
  Melbourne, VIC, Australia \\
  \texttt{guanliang.chen@monash.edu} \\
  \And
  Dragan Gasevic\\
  Faculty of Information Technology\\
  Monash University\\
  Melbourne, VIC, Australia \\
  \texttt{dragan.gasevic@monash.edu} \\
  \And
  Zachari Swiecki\\
  Faculty of Information Technology\\
  Monash University\\
  Melbourne, VIC, Australia \\
  \texttt{zach.swiecki@monash.edu} \\
}

\begin{document}

\maketitle
\begin{abstract}
  Knowledge Management is crucial for capturing and transferring expertise within universities, especially in high staff turnover contexts where expertise loss disrupts teaching. Documenting teachers' workflows is time-intensive and diverts experts from core responsibilities. Sequential Pattern Mining (SPM) leverages log data to identify expert workflows, offering an automated alternative to represent workflows but requiring transformation into intuitive formats for novice educators. This paper introduces Visual Process Representations (VPR), a design approach combining SPM, Knowledge Management processes, and storytelling techniques to convert expert log data into clear visualisations. We detail the design phases and report a study evaluating visual affordances (text lists vs. pictorial-style) and teachers' perceptions of four versions of the VPR with 160 higher teachers on Prolific. Results indicate improved task performance, usability, and engagement, particularly with enriched visuals, though process memorability and task time improvements were limited. The findings highlight VPR's potential to visualise workflows and support novice educators.
\end{abstract}


\section{INTRODUCTION}

Knowledge Management (KM) is a strategic process aimed at capturing, organising, sharing, and using knowledge to achieve specific goals \cite{Carroll2003, ishak2010integrating, Sangeeta2015}. In knowledge-intensive settings like universities, maintaining continuity in teaching practices is particularly challenging due to high staff turnover, which often leads to the loss of valuable expertise \cite{LiuPD2020, arun2005360, ravikumar2022impact}. While universities have adopted manual and digital KM systems—such as workflow documentation repositories (e.g., wikis) and electronic support tools that codify course delivery processes (e.g., content management systems) \cite{Green2022, DAGENAIS2020101778}—these approaches are often time-consuming and labour-intensive, fail to capture the nuanced knowledge of expert teachers, and distract them from their regular tasks \cite{Green2022, Fernandez-Nieto2024}. This creates an urgent need for more efficient methods to capture expert\footnote{Expert are defined as those teachers who demonstrate high expertise in performing teaching tasks and therefore do not rely on any external knowledge} teachers’ knowledge in a form that is accessible and actionable for novice\footnote{Novices are defined as those teachers who demonstrate low expertise and require additional knowledge or guidance to perform teaching tasks.}  teachers. Without such approaches, critical procedural know-how is often lost when experienced academic staff leave, forcing new staff to rely on inefficient trial-and-error strategies \cite{mcquade2007will, Meyer2006}. Addressing this challenge requires KM solutions that not only automate the capture of expert knowledge but also present it in intuitive formats that support onboarding and knowledge continuity.

Automated methods, such as Sequential Pattern Mining (SPM), offer effective alternatives for identifying and structuring complex workflows, including teaching tasks. The multifaceted KM  process (accessing, storing, sharing, and applying) provides a structure for understanding and capturing teachers' workflows \cite{Carroll2003, ishak2010integrating, Sangeeta2015}. SPM, known for analysing temporal and sequential data, has been applied in contexts such as collaborative distance learning \cite{perera2008clustering} and vehicle trajectory prediction \cite{Jamshed2024}. However, as noted by Jamshed et al. \cite{Jamshed2024}, applying these techniques to identify sequences and patterns in KM processes remains underexplored. By supporting the identification and organisation of KM processes (e.g., searching for information for Knowledge Access or annotating content for Knowledge Sharing) \cite{van2012process, he2021leveraging}, SPM offers a systematic and scalable approach to structuring workflows.

Despite the potential of SPM and KM frameworks for automating knowledge capture, a significant challenge lies in translating structured sequences or processes into intuitive and interpretable visualisations that can be effectively used by others \cite{lamghari2022process}. Novice teachers taking on new responsibilities need accessible, usable knowledge to effectively follow expert workflows, especially when additional guidance is unavailable. Existing visualisation techniques, such as Business Process Modelling Notation (BPMN) and Petri Nets, are often unsuitable for non-experts due to their reliance on specialised modelling languages and emphasis on completeness over clarity \cite{abubakre2021implementing, VERGIDIS200891}. To address this, data storytelling (DS) principles and pictorial representations offer a promising alternative by simplifying complex visualisations through narrative elements and visual cues that enhance comprehension \cite{Hongbo2024}, engagement, and recall \cite{Schulz2013, Dykes2015, Knaflic2017}. For example, storytelling principles have been successfully used to communicate multimodal sensor data in nursing contexts, helping non-experts interpret complex logs as meaningful narratives \cite{Martinez-Maldonado2020}. Building on this, data comics—a narrative form of DS—introduce techniques such as sequential art, pictorials, and digestible visual chunks to present processes in an accessible format \cite{bachCHI2018Comicdesign, InteractiveComicsWang2022}. Studies like Dospan and Khrykova \cite{Dospan2023} illustrate how manually constructed, comic-inspired workflows can support understanding, but also highlight the need for scalable, automated, and narrative-driven representations that strike a balance between usability and functional depth.


Building on KM foundations for codified knowledge—capturing knowledge in explicit and accessible forms such as documents \cite{bosua2013aligning}—, this paper addresses the challenge of capturing and sharing easy-to-interpret visual teachers' workflows by introducing a novel design approach to generate \textit{Visual Process Representations}. By combining automated methods like SPM with KM frameworks centred on knowledge processes, the proposed approach captures and structures workflows while incorporating storytelling and comic principles to visualise them in intuitive and engaging formats. This dual-pronged strategy transforms log data (e.g., clicks or keystrokes) from expert teachers into clear, interpretable, and actionable representations, making complex teachers' workflows more accessible to novice teachers.

The contributions of this paper are twofold: (i) documenting the human-centred design approach, including the exploratory phase, which identifies key design requirements and goals; the generative phase, detailing the modelling process and iterations for conceptualising Visual Process Representations; and (ii) an evaluation phase, presenting results from an evaluation with 160 higher education teachers using Prolific. The evaluation compares textual lists of steps and pictorial-style representations, with and without contextual information (e.g., screenshots indicating actions or links to repositories accessed by expert teachers). Key metrics include task performance, completion time, memorability (recall of process steps without the visual representation), and user perceptions of engagement and usability. Results indicate that Visual Process Representations effectively support novice teachers in completing teaching tasks and are perceived as engaging, particularly when enriched with visual elements. However, further research is needed to balance contextual information (e.g., see Table \ref{tab:passive_event_types}) and mitigate cognitive load, ensuring optimal usability. Teachers expressed a strong interest in adopting these representations, highlighting their potential to facilitate knowledge sharing and capture workflows in educational contexts.

\section{BACKGROUND}

\subsection{Knowledge Management in Education Settings}

Knowledge Management (KM) plays a pivotal role in capturing and sharing expertise within people, particularly in knowledge-intensive fields like education \cite{LiuPD2020}. In universities, where staff turnover is often high (e.g. retirement, moving to other parts of the institution, taking on a position with another organisation, taking over an existing course, and appointment of sessional staff and teaching assistants), documenting teachers' workflows is essential to prevent knowledge loss and ensure continuity in teaching and learning practices \cite{arun2005360,ravikumar2022impact}.


Bose and Sugumaran \cite{Bose2003} outlined KM processes that address the phases of knowledge flow, from creation to application, which help define KM and its associated activities \cite{Wong2004}. Experts engage in KM processes both tacitly and explicitly in their daily workflows \cite{DAGENAIS2020101778}. These processes (accessing, storing, sharing, and applying knowledge) offer a structured method for capturing and sharing expertise, particularly to novices. In the context of education, this is critical for mitigating turnover effects 
and enabling improved decision-making. For instance, teachers engage in activities such as navigating university repositories (\textit{Knowledge Access}), creating and uploading resources (\textit{Knowledge Storage}), annotating or commenting on materials for collaboration or personal use (\textit{Knowledge Sharing}), and applying institutional knowledge (e.g., organisational policies) to complete teaching tasks (\textit{Knowledge Application}).

Understanding KM processes provides valuable insights into what KM entails and how it can be applied to capture and share complex teaching tasks \cite{Rubenstein-Montano2015}. Drawing on these frameworks \cite{alavi2001knowledgeProcesses, andreeva2011knowledge, Carroll2003, ishak2010integrating}, which highlights the stages of access, storage, sharing, and application/reuse, this paper leverages KM processes to transform complex workflows into interpretable visual representations captured from experts' log data when using university repositories. These visual representations aim to bridge the gap between expert knowledge capture and novice understanding, facilitating more effective dissemination and application of critical teachers' workflows.

\subsection{Challenges in Knowledge Capture and Sharing}
\label{sec:challengesCaptureDiss}

\textbf{Challenges in Knowledge Capture:} Knowledge capture involves systematically collecting, documenting, and preserving valuable knowledge, insights, and expertise from individuals within an organisation \cite{Green2022, Fernandez-Nieto2024}. Capturing the tacit and implicit knowledge embedded in expert teachers' workflows remains a persistent challenge \cite{hofer2008knowledge}, yet it is essential for retaining procedural know-how within educational institutions and teaching teams \cite{LiuPD2020}, particularly to mitigate the impact of high staff turnover. Common approaches include video and audio recordings in which experts demonstrate their teaching workflows for others to observe and learn \cite{DAGENAIS2020101778, dalkir2013knowledge}. Recent KM research has also highlighted the use of live tracking technologies (e.g., Microsoft Teams, SharePoint) to capture fine-grained process changes \cite{Green2022}, as well as the assignment of knowledge management leaders who promote ongoing documentation and sharing practices \cite{Sahibzada2022}. However, these approaches have not fully explored how to meaningfully reuse knowledge captured from trace data. While such strategies support the transformation of tacit knowledge into explicit knowledge, they often rely on the expert's ability to articulate their actions—a process that is time-consuming, labour-intensive, and may divert attention from core teaching responsibilities \cite{Green2022}.

Automated methods like SPM and Process Mining (PM) identify recurrent sequences, offering insights into activity flows \cite{he2021leveraging}. SPM, when combined with modern capabilities for capturing user logs (e.g., clickstreams, navigation paths, and annotations) unobtrusively and at appropriate levels of granularity \cite{van2012process}, holds a significant potential for capturing expert knowledge in the form of workflows. Similarly, PM analyses temporal and sequential data to systematically identify, structure, and visualise workflows.

Although SPM techniques have been applied in various contexts, such as identifying knowledge-sharing patterns in collaborative distance learning \cite{perera2008clustering}, analysing students sequential behaviours \cite{Villalobos2024}, and vehicle trajectory prediction \cite{Jamshed2024}, to the best of our knowledge --and as noted in the recent literature review by Jamshed et al. \cite{Jamshed2024} -- SPM has not yet been applied to analysing KM processes as a comprehensive framework. Given its ability to analyse temporal sequences, SPM has the potential to identify patterns in accessing, storing, sharing, and applying knowledge. This capability could form the foundation for organising workflows and systematically categorising teaching activities into KM subprocesses (e.g., searching for Knowledge Access, content annotation for Knowledge Sharing) by arranging log data into logical, structured sequences.

\textbf{Challenges in Sharing -- Visualising Knowledge:} Even when workflows are effectively captured, presenting them in a format that is accessible and actionable to be used by diverse audiences, especially novice teachers, remains a significant hurdle \cite{Dospan2023}. Existing visualisation techniques, such as BPMN or Petri Nets, are well-suited for formal and structured environments but are often unsuitable for non-process experts, casual users or in teaching contexts \cite{Dospan2023, Sarshar2005}. These methods typically require specialised knowledge of process modelling languages and often prioritise completeness over clarity, making them overwhelming and less effective for sharing. This is particularly problematic for novice users, who may lack the data literacy skills needed to decode and interpret these task workflows effectively. \cite{leeVLAT2017}.

The recent work by Dospan and Khrykova \cite{Dospan2023} introduces creative approaches to process visualisation through data storytelling, particularly using data comics to depict business processes. Although this method relies heavily on manual efforts to identify workflows, create notations (e.g., BPMN), and design pictorial representations, it effectively demonstrates how visual techniques can enhance the interpretability of organisational processes. Such innovations may support novice teachers by offering step-by-step guidance enriched with contextual cues both visual and textual (e.g., screenshots or annotations), to clarify what actions to take and why. For example, a screenshot can show where an action should occur (e.g., locating a file repository), while expert annotations (e.g., notes on applying university policies) can guide users through unfamiliar tasks. However, current automated visualisation methods often lack this granularity and narrative coherence, limiting their effectiveness for knowledge sharing in educational settings.


\subsection{Data Storytelling and Data Comics for Communicating Complex Data}
\label{sec:storytelling}

Research on data storytelling (DS) highlights their potential to reduce the complexity of visualising information. DS aims to communicate complex visualisations by focusing attention to particular information through the incorporation of visual elements such as colour, lines and explanatory text to highlight key points thereby enhancing user comprehension, engagement, and recall \cite{Hongbo2024, Schulz2013, Dykes2015, Knaflic2017}. 
Similarly, data comics (DC), which is a form of DS, focus on structuring information in a sequential narrative, making it easier for users to process complex data and extract meaningful insights \cite{bachCHI2018Comicdesign, HullmanSequences}. Elements such as panels and sections break down information into manageable chunks, enhancing clarity and coherence \cite{bachCHI2018Comicdesign}. Additionally, their time- and space-oriented structure supports linear navigation and improves recall and engagement \cite{InteractiveComicsWang2022}.
By integrating storytelling principles and particular narrative elements of comics into process visualisations, our work aims to create visually engaging, and usable representations of teachers' workflows, which could potentially help teachers, especially novices, to understand complex processes.

Building on the foundations of DS \cite{Schulz2013, Dykes2015, Ryan2016, Knaflic2017, Echeverria2018, Alhadad_2018} and inspired on the DC narrative structure \cite{bachCHI2018Comicdesign, InteractiveComicsWang2022}, specific principles can be derived to design visual representations that promote easy interpretation for stakeholders. Design features for DS include: \textit{goal orientation:} ensuring the story has a clear purpose; \textit{driving audience attention:} designing visuals to focus users' attention on key ideas and data points by de-emphasising or removing unnecessary elements (decluttering); \textit{highlighting relevant data points:} using colours and visual cues to emphasise important information;  
\textit{information visualisation (InfoVis) design principles:} applying Gestalt principles, visual perception concepts, and insights from human behaviour and psychology research.

Additionally, the main DC features to improve narrative representation \cite{McCloud} in our visualisations included: \textit{sequential art and narrative flow:} using panels and space to structure the information and create coherent narrative sequences; \textit{combining words and pictures:} employing visual elements to effectively convey narratives; \textit{design patterns and panel layouts:} arranging panels to enhance navigation and readability; \textit{interaction and navigation alternatives:} allowing users to explore narratives in a non-linear fashion.

These principles guided the design of \textit{Visual Process Representations}, transforming complex workflows into engaging, interpretable, and actionable visualisations.

\subsection{Research Gaps and Contribution}

Despite growing interest in KM tools in educational contexts \cite{arun2005360, ravikumar2022impact, JARRAHIHuman-AIPartnership}, existing approaches often remain time-consuming for experts and fail to streamline knowledge sharing for novices \cite{hofer2008knowledge, Fernandez-Nieto2024, Green2022}. Empirical evidence on the effectiveness of alternative visual representations to support procedural task completion also remains limited \cite{Dospan2023, Sarshar2005}. This paper addresses these gaps by presenting a human-centred design approach to develop \textit{Visual Process Representations} that capture and illustrate teachers’ workflows. Drawing on data storytelling principles and inspired by comic-based narrative structures \cite{Martinez-Maldonado2020, Schulz2013, Dykes2015, bachCHI2018Comicdesign}, the proposed method unobtrusively captures log data (e.g., navigation logs, contextual interactions) and transforms it into structured, interpretable artefacts. Research shows that integrating storytelling elements—such as text, pictorials, panels, and layouts—enhances understanding, engagement, and long-term retention by encouraging users to interact longer with complex data \cite{InteractiveComicsWang2022, mahyar2015towards, Saket2016}. Guided by these principles, the \textit{Visual Process Representations} simplify complex workflows while preserving critical procedural details, providing actionable, memorable visualisations. Furthermore, this study offers empirical evidence on how design variations—such as textual versus pictorial styles and the inclusion of contextual information—affect teachers’ ability to interpret, complete, and recall tasks, offering practical implications for effective visual KM tool design in education.

\section{METHODS}
\subsection{Context: Teaching Tasks}
\label{sec:taskDesign}

Teachers' workflows reflect a deep understanding of the educational process—particularly task-level expertise—as a system involving interactions among various resources and individuals (e.g., teachers, researchers, and learners). These workflows typically include tasks related to accessing, creating, delivering, and updating course materials \cite{vouk1999workflow}. Their inherent complexity, contextual dependencies, and dynamic nature make them particularly difficult to document (capture) and interpret (share in usable formats)—especially for novice teachers who are required to take over responsibilities without prior exposure to the process \cite{Vouk1997, hofer2008knowledge}.

\begin{figure}[ht]
\centering
\includegraphics[width=1\textwidth]{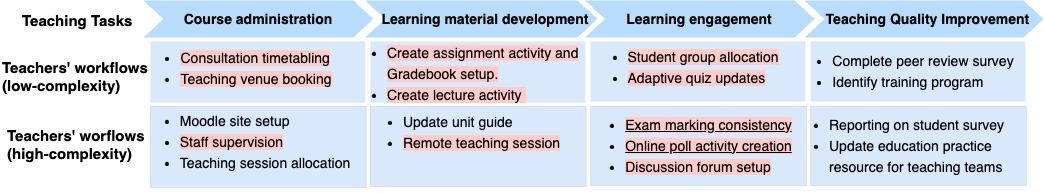}
\caption{Representative teaching tasks performed by higher education teachers in their teaching workflow. 
\textbf{Red} indicates those tasks that are performed more than once per semester. We underscore the tasks selected for evaluation in this study.} 
\label{fig:teachingtasks}
\end{figure}

\begin{figure}
    \centering
    \includegraphics[width=1\linewidth]{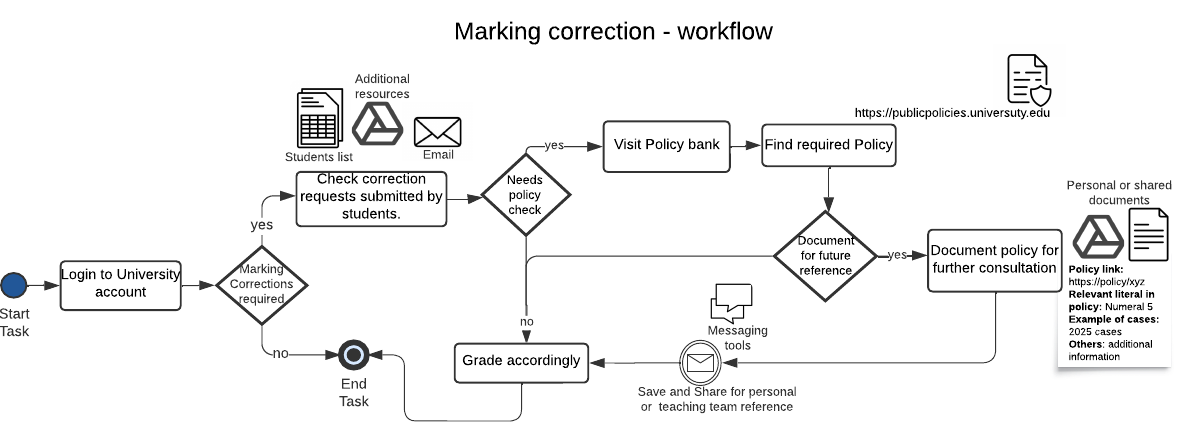}
    \caption{Teaching task workflow for making corrections on student assignments.}
    \label{fig:taskFlow}
\end{figure}

Teachers, individually or in teams, regularly carry out complex tasks to support their courses, such as creating and documenting lecture activities, ensuring consistency in exam marking, or updating teaching resources collaboratively. These tasks often involve workflows that require navigating multiple information systems and accessing a range of resources. For example, marking requires a high level of consistency—especially in team settings—to ensure fairness and uphold student outcomes. As illustrated in Figure \ref{fig:taskFlow}, the marking correction task often requires teachers to access university repositories (e.g., student lists), consult documents (e.g., policy guidelines or calendars), and store relevant information (e.g., documentation of special cases) for future individual or team reference on how to apply the policy. Without proper guidance, novice teachers taking on these responsibilities may unintentionally introduce inconsistencies. In addition, managing teaching materials or integrating activities to foster student engagement adds to the overall complexity of teaching tasks.

The inherent challenges of teaching are often intensified by handover processes that result in the loss of valuable knowledge held by expert teachers. Over time, senior educators develop deep task-level expertise—particularly procedural know-how for coordinating teaching activities, such as identifying effective strategies (e.g., creating feed-forward polls to gather student insights). When experienced teachers leave unexpectedly, novice teachers are left to navigate procedures and locate resources on their own—a task made more difficult in university settings where information is frequently dispersed across multiple systems.



Based on the co-design sessions described by Fernandez-Nieto et al. \cite{Fernandez-Nieto2024} 
and evaluations of teachers' workflows reported in Sha et al. \cite{ShareFlows2025}, 
we identified 19 representative teaching workflows across four categories: Course Administration, Learning Material Development, Learning Engagement, and Teaching Quality Improvement (see Figure \ref{fig:teachingtasks}). In these sessions, teaching teams — including Chief Examiners (CEs), Lecturers (Ls), and Tutors (Ts) \footnote{CEs are responsible for course coordination, Ls manage course delivery and coordination, and Ts support course management, marking, and tutorial delivery} — collaboratively mapped their workflows using web-based collaborative tools (e.g., Miro). Through affinity diagramming, we systematically analysed and categorised these workflows, providing insights into the nuanced practices of expert teachers and the complexity of their tasks. This understanding informed and motivated the design of \textit{Visual Process Representations}, which aim to simplify and communicate teachers' workflows for use by novice teachers. 

\subsection{Visual Process Representations Design Approach}
\label{sec:designApproach}


To capture expert knowledge embedded in teachers' workflows and make it accessible to novice educators, we propose a design approach for generating \textit{Visual Process Representations}. This approach follows a three-phase human-centred design framework \cite{evenson2010designing, ferreira2023interactions}. Originally introduced by Evenson and Dubberly \cite{evenson2010designing} in the context of \textit{Designing for Service}, this framework supports our goal of understanding teaching practices from the users’ (teachers’) perspective in order to address their needs effectively. Similar frameworks have been applied in other domains, such as communicating complex challenges in climate change \cite{ferreira2023interactions}. Our design process includes three phases: the \textbf{\textit{Exploratory phase}} (observation and reflection), informed by insights from prior co-design sessions \cite{Fernandez-Nieto2024}, 
which supported teachers in mapping their workflows and identify their needs (see Section \ref{sec:phase1}); the \textbf{\textit{Generative phase}} (ideation and iteration), where these insights were used to analyse log data and develop prototypes that incorporate storytelling and pictorial elements (see Section \ref{sec:phase2}); and the \textbf{\textit{Evaluative phase}} (testing and analysis), during which higher education teachers assessed the prototypes to inform further refinements (see Sections \ref{sec:study} and \ref{sec:results}).

\subsubsection{Phase 1: Exploratory phase}
\label{sec:phase1}
\label{sec:DR}


The work by Fernandez-Nieto et al. \cite{Fernandez-Nieto2024}  
describes a two-year study with university teachers, highlighting the need to facilitate the handover of teaching tasks. Through co-design workshops with six teaching teams (13 teachers) \cite{Fernandez-Nieto2024} 
, identified several \textbf{design requirements} (DR) to address the need for knowledge management (KM) systems supporting the transfer of teaching workflows. Teachers emphasised the need for capturing and accessing knowledge at \textit{different levels of granularity}, enabling process-level details to be captured flexibly \textbf{(DR1)}. They also highlighted the importance of capturing fine-grained information—such as notes, images, and general details—for future reference or sharing with others \textbf{(DR2)}. Additionally, mechanisms to connect teachers with expert knowledge to support retention and best practices were deemed essential \textbf{(DR3)}. Finally, participants stressed the need for timely access to relevant knowledge pieces when required \textbf{(DR4)}.


From this exploration, we derived key design goals (DGs). Given the complexity of teaching workflows, incorporating storytelling principles and inspiration from comic structures was essential. As described in Section \ref{sec:storytelling}, these elements help break down complexity into manageable steps that users can consume at their own pace \cite{Dospan2023, bachCHI2018Comicdesign}. They also support quick navigation and offer different levels of engagement—from providing an overview to enabling deeper exploration \cite{InteractiveComicsWang2022}. Our representations are automatically generated using Process Mining (PM) methods, which logically organise workflow steps. Given the complexity of these structures, the integration of storytelling principles enhances interpretability by presenting the information as \textit{Visual Process Representations}. These are delivered through panels, sections, and sequences that provide a clear narrative flow. Figure \ref{fig:StorytellingComics} illustrates the storytelling and comic elements embedded in the visual design of teachers’ workflows.

\begin{itemize}

    \item[DG1] \textbf{Narrative simplicity and goal alignment}. Visual Process Representations use storytelling principles to provide clarity and simplicity. For instance: 
    i) \textit{Goal-oriented design} clearly defines the process's purpose, guiding novice teachers to understand the task scope.  
    ii) \textit{Audience attention} is directed through visual cues, such as highlighting key steps with contrasting colours or annotations. 
    iii) \textit{Appropriate visuals} provide contextual insights with screenshots and text annotations, making the representations intuitive for novices.

    \item[DG2] \textbf{Intuitive navigation and structured flow}. Inspired by the structure of comics our Visual Process Representations ensures: 
    i) \textit{Sequentiality} with process steps and panels that are arranged in a temporal order, enabling users to follow a clear narrative. 
    ii) \textit{Detail and overview}, allowing users to switch between high-level structures and specific details through Sections (KM subprocesses). iii) \textit{Unexpected logic} where repetitions, hierarchical sequences, or exceptions are visualised to reflect the natural complexity of teaching tasks. 
    Process mining detects action patterns such as navigating resources (Knowledge Access) or annotating documents (Knowledge Sharing), which are mapped to KM subprocesses and grouped under sections for simpler navigation.
   
    \item[DG3] \textbf{Adaptable Visualisation Options}. Visual Process Representations offer adaptable presentation modes to cater to diverse user preferences: 
    i) \textit{Textual list representation} for users preferring traditional formats. 
    ii) \textit{Pictorial representations} enhance process steps with images and Panels, reflecting a comic-style narrative structure. 
    iii) \textit{Contextual data flexibility} lets users toggle between representations with or without screenshots or annotations, accommodating different user preferences and providing the appropriate level of support for novices.
\end{itemize}

\subsubsection{Phase 2: Generative Phase}
\label{sec:phase2}

Building on the needs identified by teachers, the design requirements, and the design goals that emerged from the previous phase, in this phase we aimed to conceptualise the designing of \textit{Visual Process Representations} using log data and iteratively design low-fidelity and high-fidelity prototypes. 
Thus, this phase involves two key components: (i) a modelling process that converts log data into structured visual representations, and (ii) iterative design sessions to define these representations using the design requirements and goals previously described. 

\paragraph{\textbf{(i) Modelling process: From log data to Visual Process Representations}}
\label{sec:modellingProcess}

\begin{figure}[htb]
    \centering
    \includegraphics[width=1\linewidth]{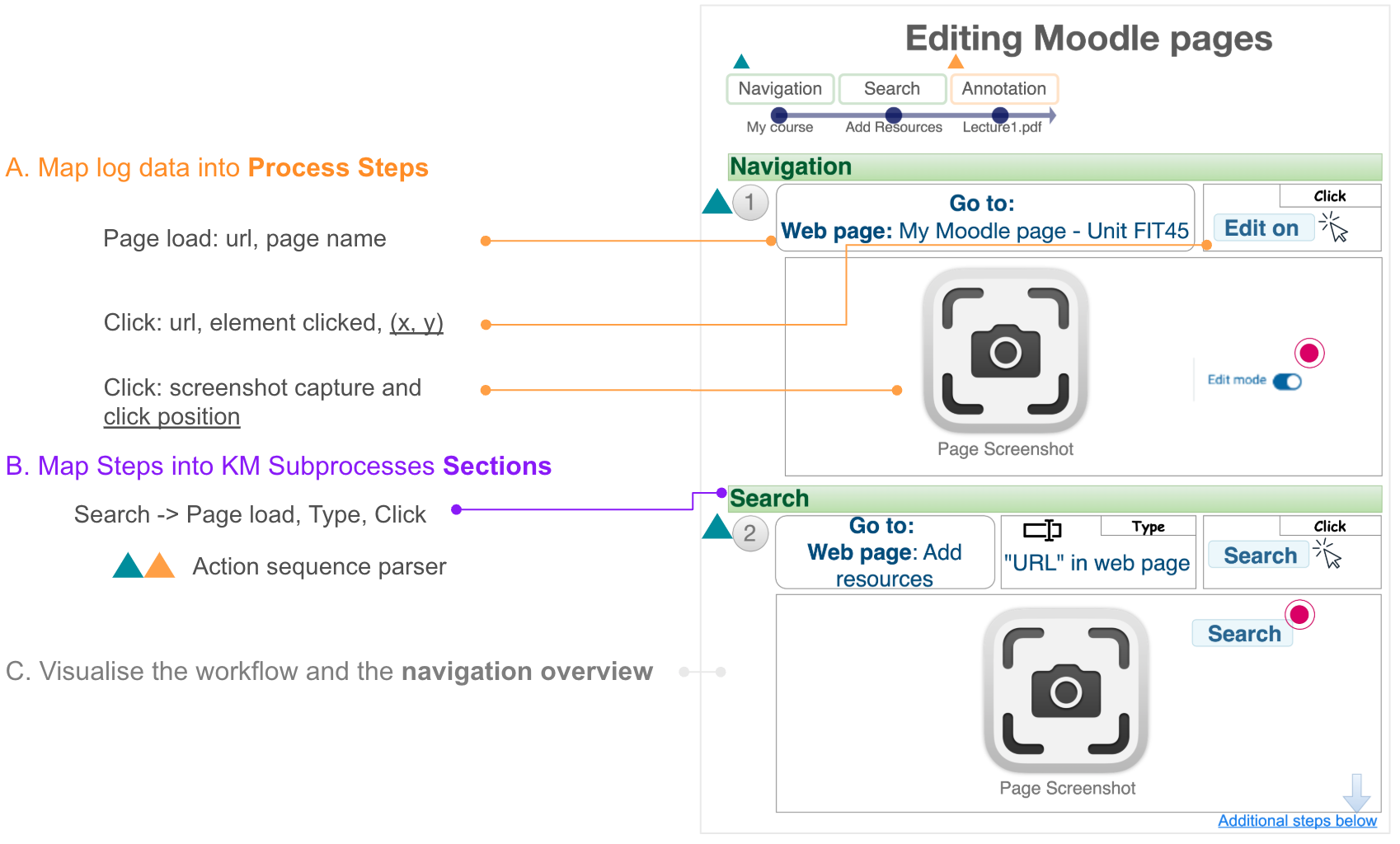}
    \caption{Modelling Process for generating Visual Process Representations: A. Log data captured from teachers' workflows are mapped to sequential process steps. B. Steps are mapped to knowledge management subprocesses and visualised as Sections or organised sequences of actions. C. The Visual Process Representation maps the teachers' workflows in an easy-to-follow sequence designed for novice teachers, while also providing a process overview to facilitate navigation.}
    \label{fig:teaser}
\end{figure}

Transforming log data into Visual Process Representations involved three stages: A) mapping log data to a sequence of process \textbf{steps} (Figure \ref{fig:teaser}--A); B) mapping task steps to knowledge management (KM) subprocesses in \textbf{sections} (Figure \ref{fig:teaser}--B); and C) \textbf{visualising} the workflow as a Visual Process Representations (Figure \ref{fig:teaser}--C). Each stage is described as follows:

\begin{itemize}
    \item [\textbf{A)}]  \textbf{Mapping log data to a sequence of process steps.} While an expert completes a task, log data are captured as process steps. The first stage involves capturing fine-grained process log data (e.g., clicks and keystrokes) along with contextual information (e.g., URLs, timestamps, and cursor positions) across the different web platforms used by teachers during their teaching tasks (e.g., Learning Management Systems, Zoom, PollEverywhere). This log data is then processed to identify significant \textit{steps} within the teaching task (Figure \ref{fig:teaser}--A), by identifying individual instances of teacher actions in the log (e.g., click to a specific URL to access a repository). The collection of these interactions is unobtrusive during expert teachers' regular workflows, via a browser extension \footnote{https://goldmind.monash.edu/
    }. 
    The list of captured events their descriptions and contextual information is in Table \ref{tab:passive_event_types}.  

\begin{table}[ht]
\caption{Log data type and additional information used to define steps in the process.}
\resizebox{\textwidth}{!}{%
\begin{tabular}{p{0.1\textwidth}p{0.4\textwidth}p{0.5\textwidth}} \toprule
\textbf{Type} & \textbf{Description} & \textbf{Additional Information} \\ \midrule
Click & Clicking on a specific element within a web page. & Element type (e.g., button), clicking coordinates (x, y), texts within clicked elements (if available), page URL, timestamps. \\ \midrule
Keyup & Keyboard activity within a web page. & Key value typed, page URL, timestamp \\ \midrule
Select & Selecting a specific element within a web page. & Selected span of text, page URL, timestamp \\ \midrule
Scroll & Scrolling activity within a web page. & Scrolling distances, page URL, timestamp \\ \midrule
Switch-tab & Switching to a different tab. & Page URL, timestamp \\ \midrule
Focus & Mouse focusing (e.g., on an input field or iframe). & Element name, type, page URL, timestamp \\ \midrule
Change & Changing activity within a web page (e.g., check a checkbox) & Element name, type, new value (e.g., updated text), page URL, timestamp \\ \midrule
Submit & Submitting activity within a web page. & user ID, page URL, timestamp \\ \midrule
Navigate & Navigating activity to another web page. & Page URL \\ \midrule
Close & Closing a web page. & Element name, type, page URL \\ \bottomrule
\end{tabular}%
}

\label{tab:passive_event_types}
\end{table}

\item[\textbf{B)}] \textbf{Task steps into KM subprocesses - Sections.} In this stage, we use SPM to identify potential sequences of task steps associated with KM processes. These steps are grouped into meaningful Sections based on KM processes, with additional contextual information (e.g., screenshots or annotations) provided where necessary (Figure \ref{fig:teaser}--B). To code process steps into KM processes, we refer to established KM frameworks \cite{andreeva2011knowledge, Carroll2003, ishak2010integrating, Sangeeta2015}, which outline four KM processes (Access, Store, Share, and Apply).

In the context of teaching tasks, these major processes were further decomposed into KM processes and subprocesses (see table \ref{table:KMProcesses}).  \textit{Knowledge Access} includes subprocesses where teachers navigate university resources or repositories to locate or search for materials. \textit{Knowledge Storage} involves teaching actions where teachers upload, fill in information, or add resources to university repositories (e.g., Moodle). \textit{Knowledge Sharing} encompasses actions such as highlighting and annotating resources, whether for personal use or to share with colleagues or students. \textit{Knowledge Application} refers to using existing knowledge when needed; completing a teaching task with a Visual Process Representation is one example.

\begin{table}[htb]
    \centering
    \caption{Main KM Processes and their subprocesses with definitions.}
    \resizebox{\textwidth}{!}{%
    \begin{tabular}{|p{0.2\textwidth}|p{0.22\textwidth}|p{0.46\textwidth}|}
        \hline
        \textbf{Main KM Process} & \textbf{KM Subprocess} & \textbf{Definition} \\
        \hline
        \multirow{2}{=}{Knowledge Access} 
        & Navigation & A KM subprocess where users explore resources. \\
        \cline{2-3}
        & Search & A KM subprocess where users search for existing resources to support their tasks. \\
        \hline
        \multirow{2}{=}{Knowledge Store} 
        & Filling information & Users capturing information while completing teaching tasks. \\
        \cline{2-3}
        & Uploading KM resources & Users uploading new teaching material or resources to the knowledge repository. \\
        \hline
        \multirow{2}{=}{Knowledge Sharing} 
        & Document annotation & Users annotate documents to share best practices or case studies for use by others or for their own reference. \\
        \cline{2-3}
        & Highlight information & Users highlight information in existing resources visible to other teachers. \\
        \hline
        \multirow{2}{=}{Knowledge Application} 
        & Interact with and use available resources & Users engage with instructional resources, such as video tutorials, in a task context. \\
        \cline{2-3}
        & Rely on automated system recommendations & Users rely on automated recommendations to highlight relevant information for teaching tasks. \\
        \hline
    \end{tabular}%
    }
    \label{table:KMProcesses}
\end{table}

\begin{figure}[!th]
    \centering
    \includegraphics[width=0.8\linewidth]{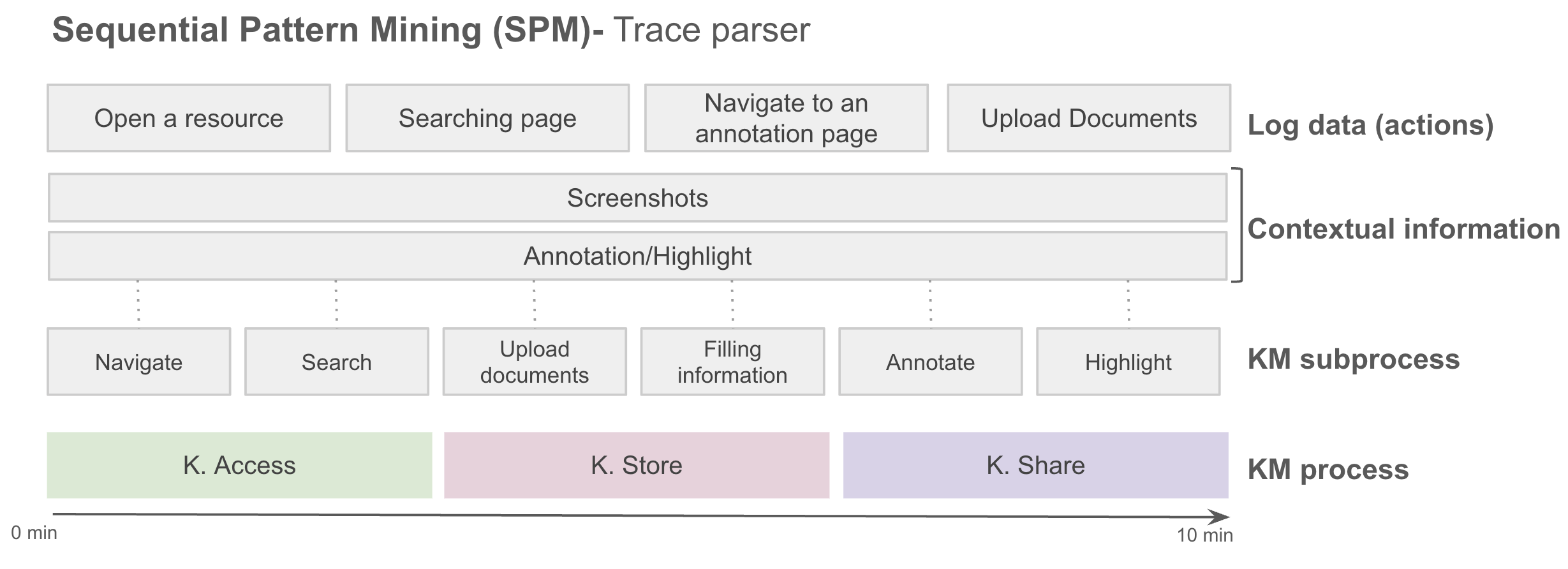}
    \caption{Mapping from log data to KM subprocesses and processes.}
    \label{fig:mapTraceSubProcesses}
\end{figure}

The transformation of task steps into KM subprocesses involves mapping these steps into sequences that reflect the temporal order of actions in the log data, ensuring logical flow and coherence in the subsequent visualisation stage. Figure \ref{fig:mapTraceSubProcesses} summarises the mapping process from navigation log and contextual data to KM subprocesses. First, log data were translated into KM teaching steps (e.g., a teacher navigating university repositories could indicate a SEARCH or NAVIGATE sequence). Then these teaching steps 
were mapped to the corresponding KM subprocesses outlined in Table \ref{table:KMProcesses}. For example, when an expert teacher highlighted a portion of text or made annotations while reading teaching policies, which are publicly accessible to other teachers, the sequence of actions was labelled as NAVIGATE<->ANNOTATE and subsequently mapped to the Knowledge Share process, as the teacher was orienting towards capturing recommendations or documentation for future public use. Actions that could not be mapped to any proposed subprocess were labelled as No\_Process.

Variations across subprocesses were identified using process variation analysis \footnote{implemented with the pm4py Python library}. Detailed variations identified in the teaching domain are included in the supplementary material. The outcome of this stage is a structured representation that preserves the richness of expert workflows (steps) and provides a high-level interpretation by grouping steps into KM subprocesses, represented as Sections in the visualisation.

\item[\textbf{C)}] \textbf{Process visualisation.} The last stage of the transformation process is the visualisation of process \textit{steps} and \textit{sections} into a Visual Process Representation. In this stage, each distinct process step is visualised within sections, while the collective sections (KM subprocesses) form the complete Visual Process Representation (Figure \ref{fig:teaser}--C). The representation includes interactive elements such as clickable URLs for accessing relevant resources (e.g., document links) and additional visual context data such as screenshots that are zoomable for detailed insights into the expert's process. These features aim to provide novice teachers with a detailed yet interpretable visual representation of a complex teaching process. Figure \ref{fig:VPR-Education} illustrates a Visual Process Representation designed to support novice teachers with KM in the context of marking corrections (described in Figure \ref{fig:taskFlow}). The Visual Process Representation visualises an expert teacher’s workflow across four main sections: i) navigating to student lists, ii) searching institutional repositories for relevant policies, iii) accessing feedback procedures, iv) and annotating key documents and multiple  steps. These visual representations were design in two iterations.
\end{itemize}

\begin{figure}
    \centering
    \includegraphics[width=1\linewidth]{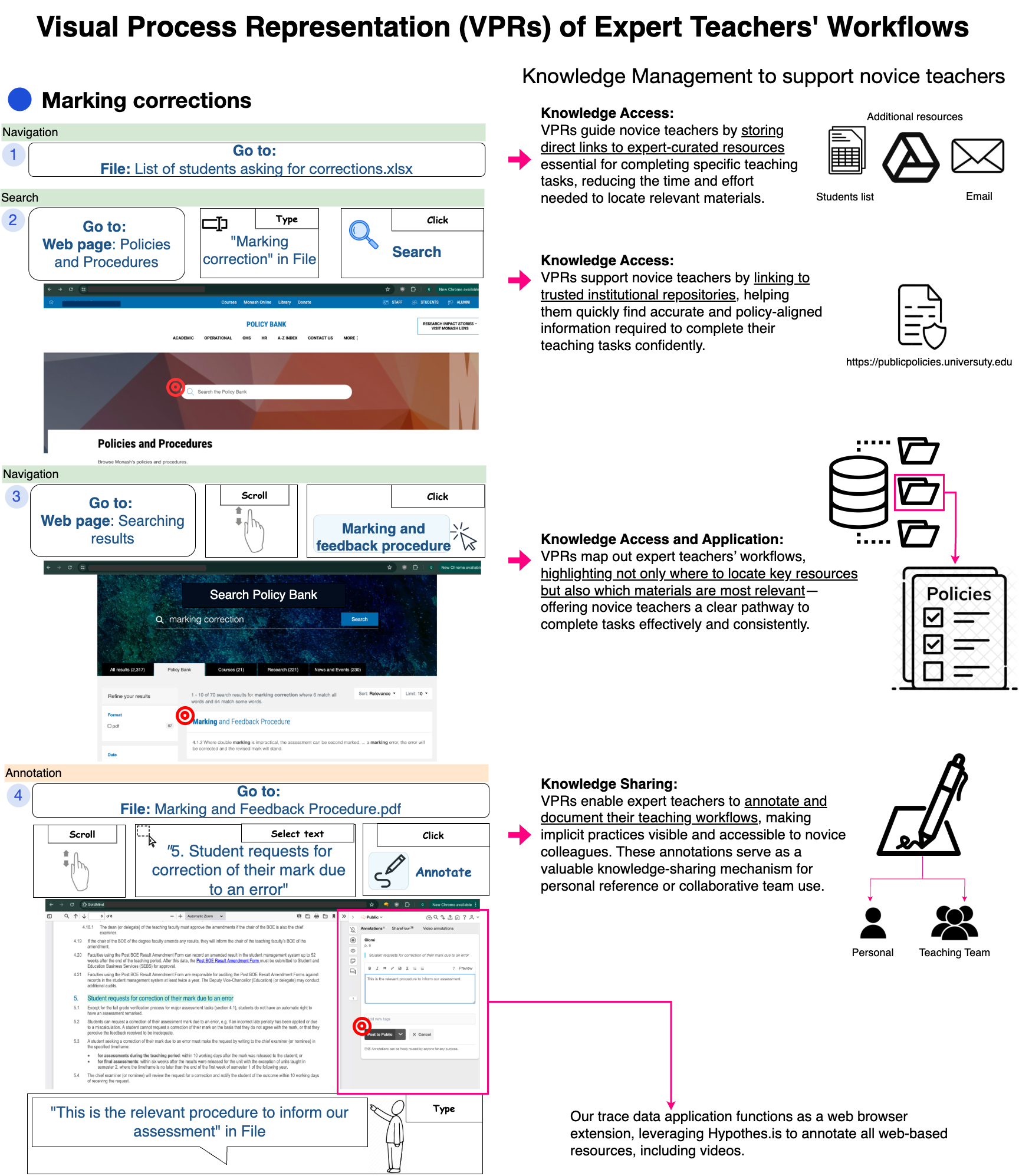}
    \caption{Example of a Visual Process Representation (VPR) for the marking correction task, illustrating an expert teacher's workflow and how the visualisation supports novice teachers in their knowledge management tasks.}
    \label{fig:VPR-Education}
\end{figure}

\begin{figure}[ht]
    \centering
    \includegraphics[width=0.6\linewidth]{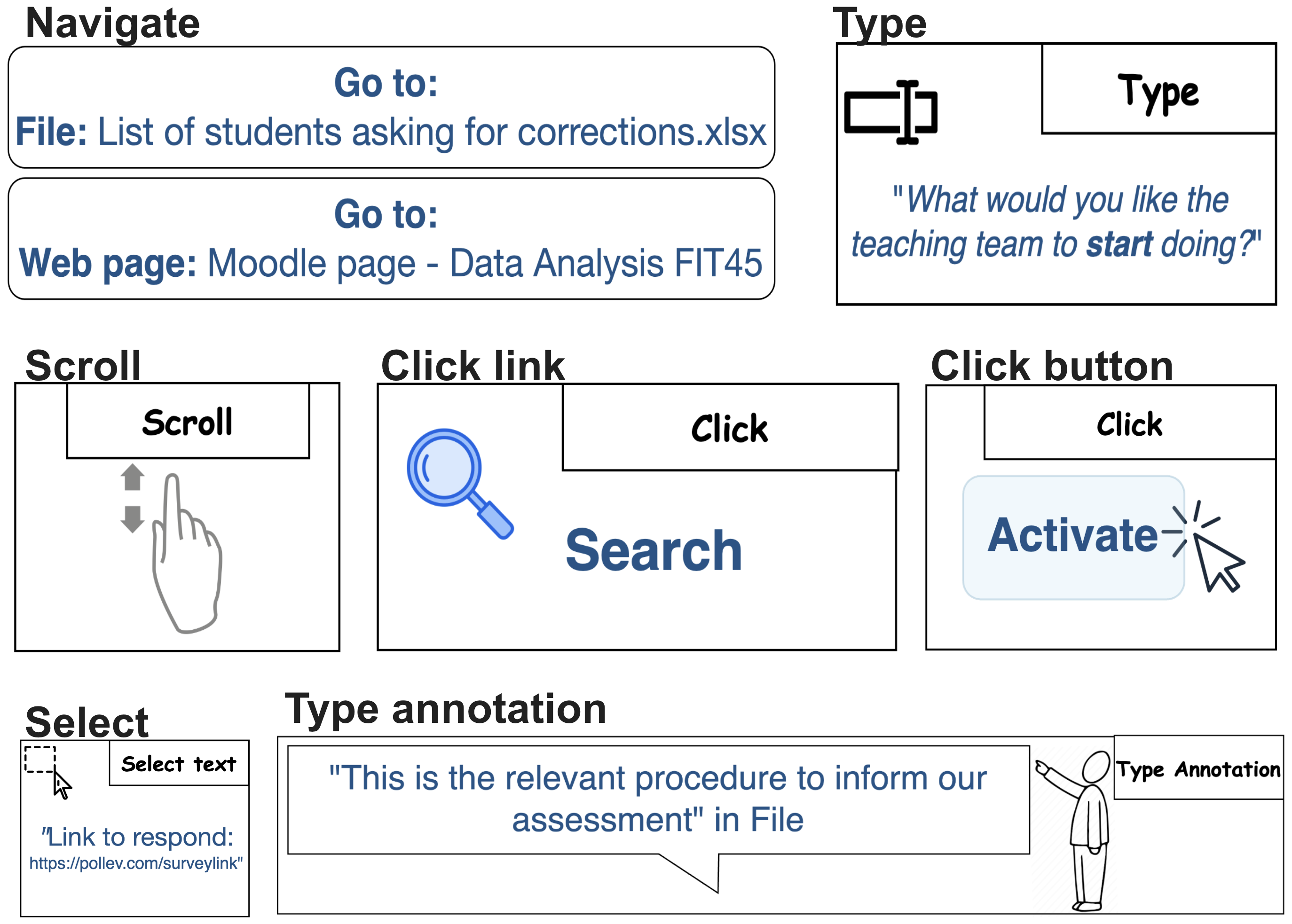}
    \caption{Dictionary of pictographic elements to map actions into Panels}
    \label{fig:PictographicPanels}
\end{figure}

\paragraph{\textbf{(ii) Prototyping Iterations}} 
The first iteration focused on creating a low-fidelity prototype that incorporated DR (1-4) (Section \ref{sec:DR}) and DG (1-3) to simplify visual complexity. The second iteration refined the prototype into a high-fidelity version, focusing on developing navigation options to improve usability.

\begin{figure}[hb]
    \centering
    \includegraphics[width=1\linewidth]{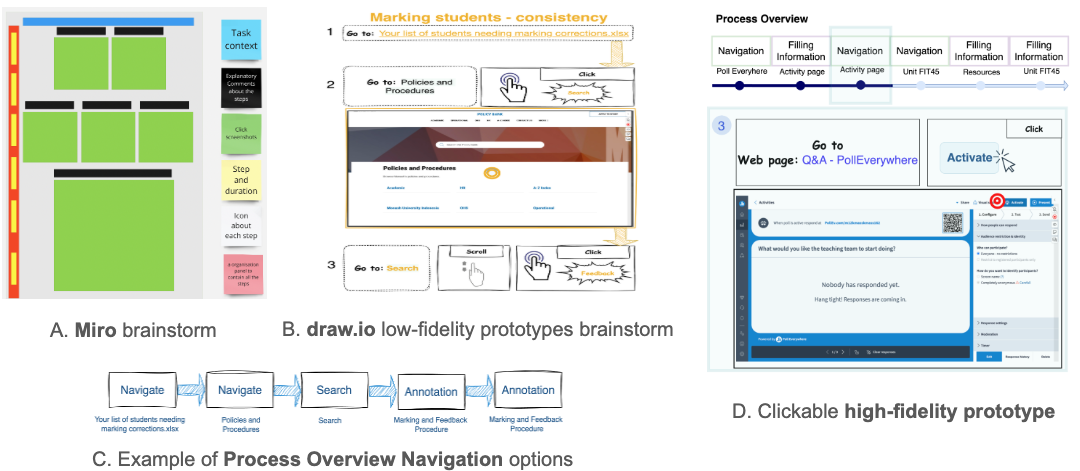}
    \caption{First and Second Iterations: A and B show the initial designs (low-fidelity prototypes). C presents one of the process overview designs. D displays a clickable high-fidelity prototype.}
    \label{fig:first-second-iterations}
\end{figure}

\paragraph{\textbf{Iteration 1: Low-fidelity prototype}}. 
Three authors, experts on human-computer interaction -HCI (average years of HCI experience = 5.66), 
worked on preliminary design ideas using low-fidelity prototypes drafted on a collaborative web tool (e.g., Miro, see Figure \ref{fig:first-second-iterations}--A). 

\begin{figure}[ht]
\centering
\includegraphics[width=.8\textwidth]{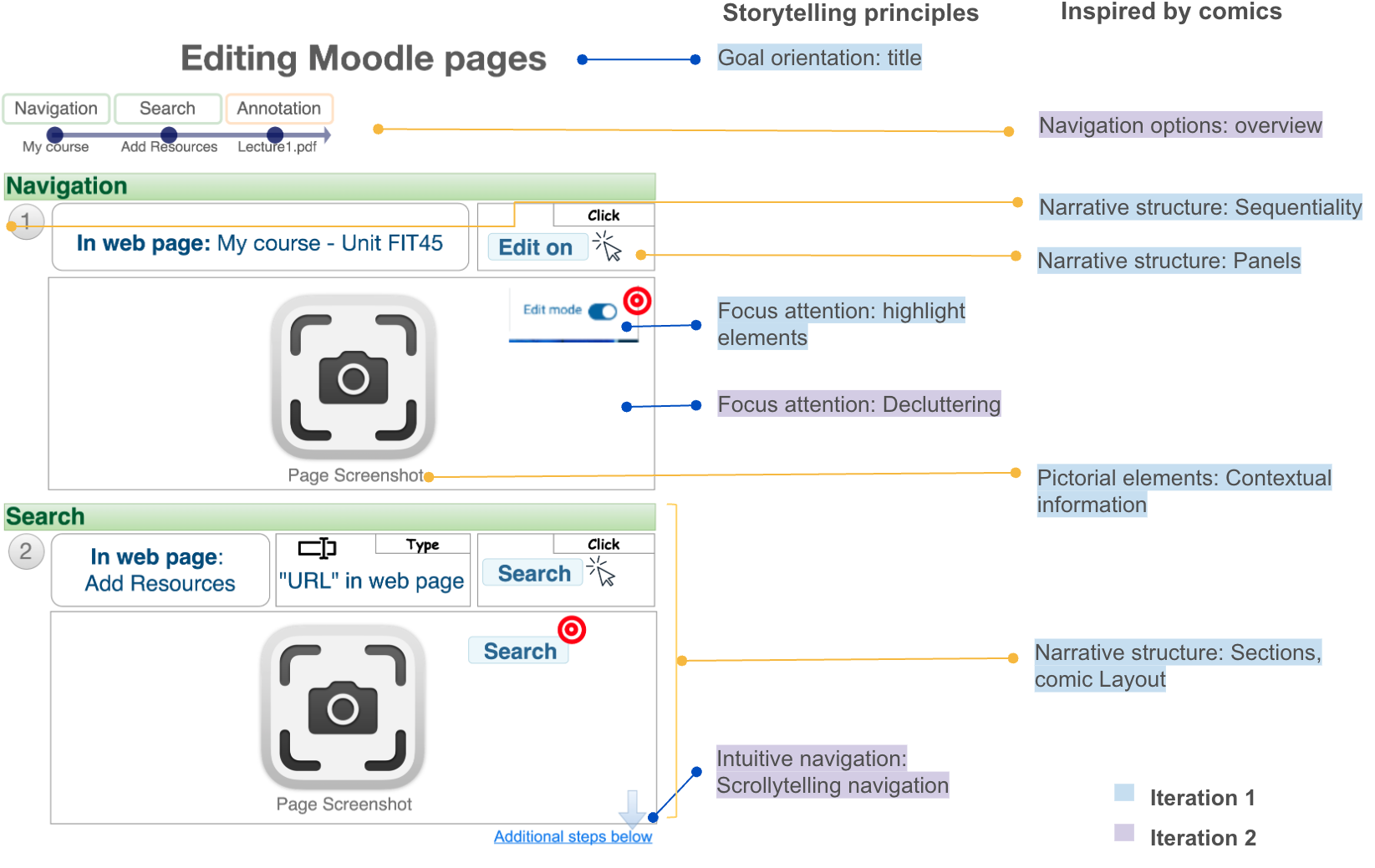}
\caption{Data storytelling and elements in the design inspired by comic-narrative incorporated into our process-pictorial approach} 
\label{fig:StorytellingComics}
\end{figure}


The primary focus of the initial iteration was to inform the visual representation, generated in the modelling process explained in Section \ref{sec:modellingProcess}, using the established  DR and DG to incorporate visualisation techniques to effectively communicate complex data. To address \textbf{DR1} and \textbf{DR2}, experts identified the potential of log data captured from regular teaching workflows as a valuable resource for building the prototype. The captured data included teachers' log data (e.g., clicks and keystrokes), and contextual information (e.g., URLs, timestamps, cursor positions, screenshots, and annotations). 
The concept of mapping log data into process steps and grouping them into sections emerged from the need to design a representation that uses captured data to facilitate knowledge sharing for both expert and novice users. In line with requirements \textbf{DR3} and \textbf{DR4}, the experts proposed the first version of the Visual Process Representations, using log data to plot process steps and sections in a sequence determined by the chronological order of the captured data. The right-hand side of Figure \ref{fig:VPR-Education} highlights how Visual Process Representation enhance knowledge access, application, and sharing by linking (using URLs) to expert-curated resources, identifying key materials, and enabling annotation for team use.

Experts incorporated storytelling principles into the Visual Process Representations. Figure \ref{fig:StorytellingComics} illustrates the design elements considered during the first iteration (highlighted in blue). In alignment with \textbf{DG1}, storytelling elements included: (i) a clear visualisation purpose, indicated by a descriptive title, and (ii) audience attention, supported by visual cues such as highlighting critical decision points in red. For \textbf{DG2}, comic-style narrative elements were integrated, including sequentiality, Panels (to represent steps), Sections, and pictorial elements such as contextual data (e.g., screenshots) to guide novice teachers through teaching tasks. A dictionary of pictographic elements was created to map each step captured in the log data (see Figure \ref{fig:PictographicPanels}). 

\begin{figure}[ht]
    \centering
    \includegraphics[width=0.8\linewidth]{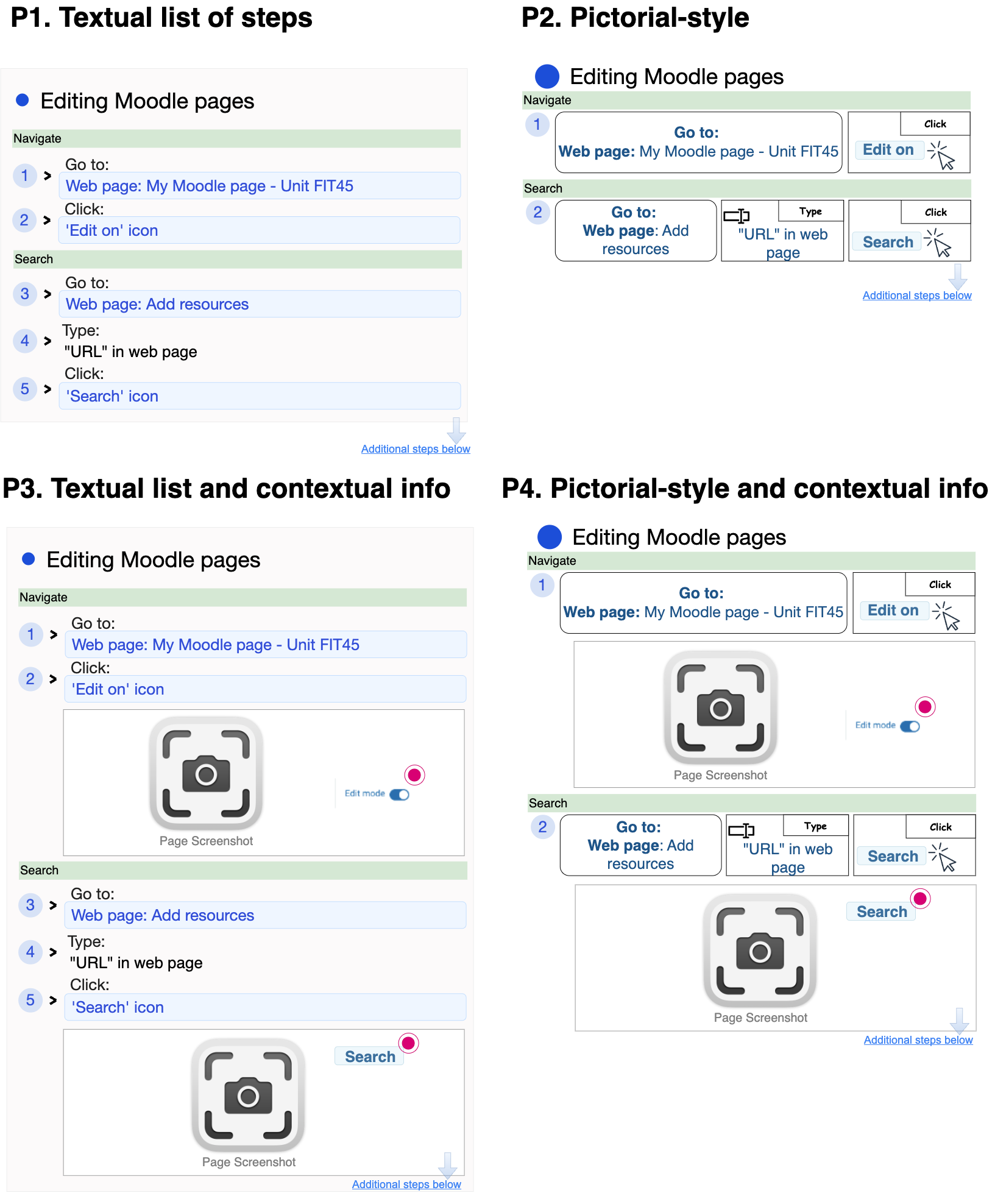}
    \caption{The study evaluated four prototypes (P1–P4), each offering a distinct representation of the KM process. P1 presented a simple textual list of steps. P2 featured a pictorial-style format, incorporating panels and icons. P3 combined a list of steps with additional contextual information. P4 integrated a pictorial-style with contextual information, offering a more visual and enriched representation.}
    \label{fig:fourPrototypes}
\end{figure}

Following \textbf{(DG3)}, novice teachers can choose the format that best supports their needs for interpreting and navigating complex teaching tasks: (i) \textit{textual lists of steps} or (ii) \textit{pictorial-style representations}. The textual alternative presents steps and KM subprocesses in plain text, offering a straightforward and familiar way to interpret workflows (see P1 and P3 in Figure \ref{fig:fourPrototypes}). The pictorial-style alternative draws on comic-narrative structures (see P2 and P4 in Figure \ref{fig:fourPrototypes}), breaking information into smaller, manageable \textit{Panels} \cite{InteractiveComicsWang2022}. Each Panel depicts an individual step, enriched with images to support a linear narrative progression and enhance interpretability \cite{bachCHI2018Comicdesign}. Sections group related Panels to visualise KM subprocesses within a teaching task, aligning with the goal of creating interpretable and actionable representations.

Contextual data (e.g., screenshots) were integrated into the prototypes to provide additional reference points and visual cues. For instance, highlighting where specific actions (e.g., a click) occur helps novices validate their task steps (see Figure \ref{fig:StorytellingComics}, red highlighting). The prototypes were designed flexibly, allowing contextual data to be included or excluded depending on user needs, ensuring accessibility and adaptability across diverse novice users.


Finally, the first author created the resulting prototype of this iteration using the \textit{drawio} tool (see Figure \ref{fig:first-second-iterations}--B). 



\paragraph{\textbf{Iteration 2: High-fidelity prototype
}} This iteration aimed to validate the modelling process, readability and navigability of the visual representation. For this purpose, two HCI experts (average years of HCI experience = 5) and two senior and one casual Lecturers from a higher education institution (average years of teaching experience = 7.5) refined the prototype multiple times using the web tool \textit{drawio}. Figure \ref{fig:StorytellingComics} illustrates the elements designed in the second iteration (shown in purple). 

A key focus was developing alternative navigation options for the Visual Process Representations (Figure \ref{fig:first-second-iterations}--C), inspired by \textbf{DG2}, to provide both detailed and overview perspectives. An overview using scrollytelling was implemented, where webpage elements move or change as teachers scroll, enhancing engagement and readability \cite{Scrollytelling2023}. Iterative refinements removed unnecessary elements to declutter the interface (\textbf{DG1}). While senior and casual teachers completed tasks, their trace data was collected via a browser extension\footnote{https://goldmind.monash.edu/query
}. The final prototype, created using HTML, CSS, and JavaScript to visualise expert teachers' data, is shown in Figure \ref{fig:first-second-iterations}--D, featuring the overview navigation at the top and the detailed process Sections highlighted.

Finally, our human-centred design approach also incorporates phase 3: \textbf{Evaluation Phase}, including testing and analysis, reported in Sections \ref{sec:study} and \ref{sec:results}. 

\section{EVALUATION STUDY}
\label{sec:study}


The motivation of this study is to evaluate alternatives in KM process visualisations (textual vs. pictorial) and the role of contextual elements. To this end, we conducted a between-subjects study using four prototypes (Figure \ref{fig:fourPrototypes}): P1) textual list of steps, P2) pictorial style, P3) textual list with contextual information, and P4) pictorial style with contextual information. This study addressed the following research questions (RQs):
\begin{itemize}
    \item[\textbf{RQ1:}] How do the different affordances of Visual Process Representations (textual list vs. pictorial, with or without contextual information) affect teachers' ability to interpret information and complete tasks in terms of accuracy, completion time, and memorability? 
    
    \item[\textbf{RQ2:}] What are the differences in teachers' perceptions of usability and engagement across the four Visual Process Representations?

    \item[\textbf{RQ3:}] What improvements do teachers recommend for each Visual Process Representation prototype, and how do these suggestions align with identified engagement and usability challenges?
\end{itemize}

\subsection{Participants}
To address our RQs and evaluate whether Visual Process Representations support teachers' workflows across varying roles and experiences, we recruited 160 higher education teachers (T1–T160) via Prolific\footnote{\href{https://www.prolific.com/}{Prolific platform}}. A power analysis recommended 40 participants per prototype. Eligibility was confirmed using a customised screener: \textit{Employment Role} ("Teacher" or "Education and Training") and \textit{Industry} ("College, University, and Adult Education"). Participant demographics are summarised in Table \ref{tab:participant_demographics}. Ethics approval was obtained 
\footnote{Our study is approved by the 
University ethics committee under number MUHREC-30388}, and all participants provided informed consent after receiving an explanatory statement outlining the study objectives and procedures.

\begin{table}[ht]
\centering
\caption{Participant Demographics by Prototype}
\label{tab:participant_demographics}
\begin{tabular}{|c|c|c|c|}
\hline
\textbf{Prototype} & \textbf{Gender} & \textbf{Ages} & \textbf{Years of Teaching Experience} \\ \hline
P1 & Female = 27, Male = 13 & 23 - 66 (M = 36, std = 12.3) & 4 - 8 (M = 7, std = 1.3) \\ \hline
P2 & Female = 28, Male = 12 & 23 - 65 (M = 31, std = 10.8) & 4 - 8 (M = 6, std = 1.3) \\ \hline
P3 & Female = 24, Male = 16 & 21 - 58 (M = 36, std = 9.8) & 5 - 8 (M = 7, std = 1.1) \\ \hline
P4 & Female = 22, Male = 18 & 21 - 26 (M = 31, std = 9.7) & 4 - 8 (M = 6, std = 1.2) \\ \hline
\end{tabular}
\end{table}


\subsection{Task selection}

To evaluate the prototypes, we selected tasks that met three criteria: 1) independently performed without assistance or feedback, 2) involving the university's teaching and learning management system (e.g., Moodle), and 3) frequently performed, ensuring relevance to teachers' daily workflows. Based on these criteria, we selected two tasks\footnote{No more than two tasks were included due to time constraints, as participants needed to perform both tasks during the session.}: Task 1) exam marking to ensure consistency (\textit{Documenting Marking}, see Figure \ref{fig:taskFlow}); and Task 2) \textit{online poll} activity creation.


\subsection{Procedure}
\label{sec:procedure}

\begin{figure}[h]
    \centering
    \includegraphics[width=1\linewidth]{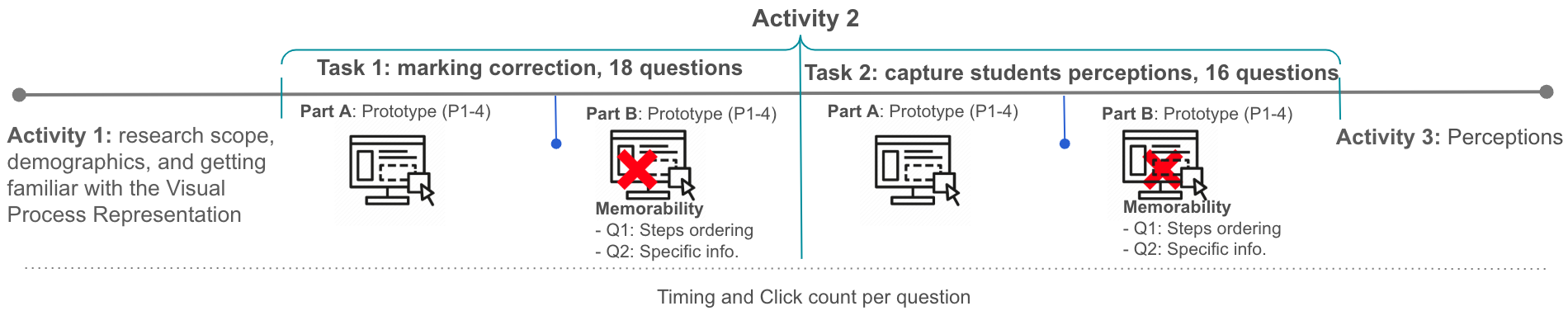}
    \caption{Study Procedure}
    \label{fig:procedure}
\end{figure}

Each participant evaluated one of the four prototypes. The prototypes represented two of the \textit{teachers' workflows} described in Section \ref{sec:taskDesign} and included all process steps and sections detailed in Section \ref{sec:designApproach}. The study was conducted using Qualtrics, with Visual Process Representations embedded within each survey question. The 1-hour evaluation consisted of three activities (see Figure \ref{fig:procedure}), outlined below:

\begin{itemize}

    \item \textit{Activity  1}: Participants provided demographic and teaching information, including courses taught, years of experience, and teaching roles (e.g., lecturer). They were also introduced to the study’s scope and the purpose of Visual Process Representations, with time allocated to familiarise themselves with the prototypes' user interfaces and navigation.

    \item \textit{Activity 2 -} \textbf{(RQ1)}:  In this main activity, participants completed two teaching tasks. 
    Task 1 consisted of 18 multiple-choice questions, and Task 2 had 16 multiple-choice questions (included in the supplementary material), designed to assess participants' comprehension of the prototypes. For example, questions asked participants to identify relevant steps, sections, or information (e.g., \textit{Which policy file needs to be accessed to complete this task?}). Each task was divided into two parts, as shown in Figure \ref{fig:procedure}: In \textit{Part A}, participants answered questions using the interactive prototype embedded in the online survey. In \textit{Part B}, the prototype was removed to evaluate memorability. We collected the correctness of each question (accuracy) and the time that participants spent answering each of the questions (completion time).
    
    \item \textit{Activity 3 -} \textbf{(RQ2-RQ3)}: We adapted the user engagement survey from O’Brien and Toms \cite{Heather2010} and a usability survey focused on visualisation evaluation \cite{Saket2016} to provide quantitative insights into participants' perceptions. Using a Likert scale (1–5), participants were asked to rate their perceptions of the usability and engagement of the prototype they interacted with. Additionally, they were asked to respond to an open question indicating whether they encountered any challenges while navigating the prototype, and were invited to provide suggestions for improvement, if applicable.
    
\end{itemize}

\subsection{Analysis}

 
To address \textbf{RQ1}, we excluded four participants from Task 1 and eleven participants from Task 2 who spent less than 30 seconds per question, as this time frame is unlikely to produce reliable responses \cite{Douglas2023}. For the remaining participants, we calculated total task scores for Parts A and B separately to assess \textbf{task accuracy} with and without the use of the prototype. 

To compare the prototypes, we conducted regression analyses using binary indicators to represent prototype pairs (e.g., P1 vs. P3). The dependent variables were (i) task scores for Part A and Part B separately and (ii) the total time on task. No covariates were included in the models. Regression coefficients indicated the magnitude and direction of the differences, while Bonferroni-corrected p-values and effect sizes were used to control for potential Type I errors resulting from multiple comparisons \cite{Armstrong2014}.

In addition, for \textbf{completion time}, we calculated the average and total time spent on each task, both with the prototype (Part A) and without it (Part B), as well as the total time spent answering all questions. A correlation analysis was performed to examine the relationship between \textit{time on task} and \textit{scores} for each prototype. Joint plots were created to visualise the time on task versus the scores, along with correlation heatmaps to summarise these relationships.

To address \textbf{RQ2}, we statistically analysed the Likert-scale responses and examined differences among prototypes in terms of usability and engagement. We also performed a regression analysis to evaluate pairs of prototypes, using the score per question.


Finally, to address \textbf{RQ3}, we analysed the 160 open-ended responses collected as described in Section \ref{sec:procedure}. 
The lead author applied an inductive approach to identify emerging themes (Table \ref{tab:issues_summary}) outlining potential enhancements for each prototype \cite{ThematicAnalysis2006}. Two researchers with HCI backgrounds then coded the data. They first conducted coding training on 30\% of the responses, followed by several discussion sessions to resolve disagreements, particularly where quotes overlapped codes. These sessions resulted in an agreement rate of 80\% on the training data \cite{MatthewQA2019}. The researchers subsequently coded the remaining responses independently, after which the coded statements were collaboratively reviewed to select representative examples highlighting opportunities for prototype improvement.

\section{Results}
\label{sec:results}

\textbf{RQ1.} \textit{How do the different affordances of Visual Process Representations affect teachers' ability to interpret information and complete tasks in terms of accuracy, completion time, and memorability}

\begin{figure}[t]
    \centering
    \includegraphics[width=1\linewidth]{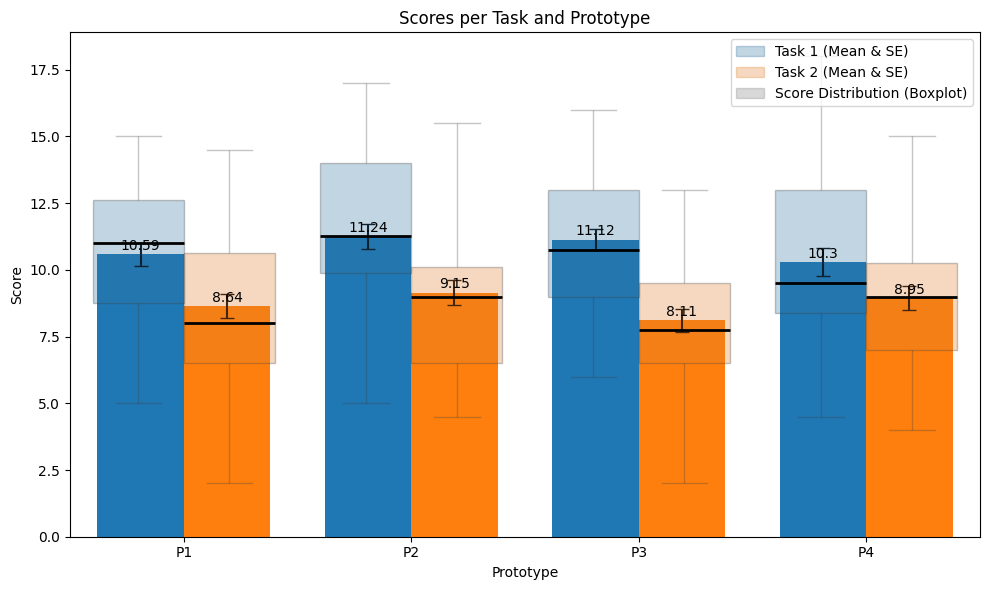}
    \caption{Score comparison across the four prototypes: Task 1 (blue) and Task 2 (orange)}
    \label{fig:prototypes-score}
\end{figure}

\textbf{Task accuracy:} Figure \ref{fig:prototypes-score} summarises the average scores per prototype and task (Part A and Part B combined). P1 and P3 represent the Visual Process Representation with textual list of steps, with only P3 including contextual information (e.g., screenshots). P2 and P4 are the pictorial-style representations of the process, and P4 also includes contextual information. From the results, we observed that all prototypes performed similarly for a given task, indicating their effectiveness in communicating information to teachers to complete tasks. In particular, P2 demonstrated a slight advantage in both tasks. In addition, the results highlight that the highest scores were achieved only under the conditions P2 and P4, reaching 18 points on Task 1 and 16 points on Task 2.

\begin{table}[ht]
\centering
\caption{Task 1. Comparison Results for parts A and B}
\renewcommand{\arraystretch}{1.2} 
\setlength{\tabcolsep}{4pt} 
\begin{tabularx}{\textwidth}{@{}lccccc|ccccc@{}}
\toprule
 & \multicolumn{5}{c}{\textbf{PART A}} & \multicolumn{5}{c}{\textbf{PART B}} \\ 
\cmidrule(lr){2-6} \cmidrule(lr){7-11}
\textbf{Pair} & \textbf{Coef} & \textbf{t} & \textbf{P>|t|} & \textbf{Corrected p} & \textbf{Cohen's d} & \textbf{Coef} & \textbf{t} & \textbf{P>|t|} & \textbf{Corrected p} & \textbf{Cohen's d} \\ 
\midrule
P1 vs P2 & 0.7066 & 1.955 & 0.054 & 0.324 & -0.442 & 0.1592 & 0.336 & 0.738 & 1 & -0.076 \\ 
P1 vs P3 & 0.5816 & 1.641 & 0.105 & 0.63 & -0.371 & -0.0158 & -0.037 & 0.971 & 1 & 0.008 \\ 
P1 vs P4 & -0.0526 & -0.128 & 0.899 & 1 & 0.029 & -0.2632 & -0.568 & 0.572 & 1 & 0.130 \\ 
P2 vs P3 & -0.1250 & -0.339 & 0.736 & 1 & 0.075 & -0.1750 & -0.407 & 0.685 & 1 & 0.091 \\ 
P2 vs P4 & -0.7592 & -1.799 & 0.076 & 0.456 & 0.407 & -0.4224 & -0.906 & 0.368 & 1 & 0.205 \\ 
P3 vs P4 & -0.6342 & -1.524 & 0.132 & 0.134 & 0.345 & -0.2474 & -0.592 & 0.555 & 1 & 0.792 \\ 
\bottomrule
\label{table:pairwiseTask1}
\end{tabularx}
\end{table}

\textit{Task 1:} Table \ref{table:pairwiseTask1} summarises the comparison of coefficients, t-values, and p-values between prototypes for Task 1, with key insights highlighted in blue. Comparing Prototypes 1 (textual list) and 2 (pictorial-style view without contextual data), regression analysis indicates an improvement in Task 1 accuracy by 0.7066 points for Prototype 2 (t = 1.956, p = 0.324, d = -0.442). While not statistically significant, this suggests Prototype 2 may enhance performance over Prototype 1.

For Prototypes 1 and 3 (textual lists, with Prototype 3 including contextual data), results show a positive coefficient for Prototype 3, improving performance by 0.581 points (t = 1.64, p = 0.63, d = -0.371). The inclusion of contextual data, such as screenshots, likely aided task completion. Conversely, comparing Prototypes 2 and 4 (both pictorial-based, with Prototype 4 adding contextual data), Prototype 2 outperformed Prototype 4 (t = -1.79, p = 0.456, d = 0.407). Although not statistically significant, this suggests Prototype 2 may be more intuitive for Task 1.

\textit{Task 2:} Table \ref{table:task2PairsStatisticalAnalysis} summarises the coefficients, t-values, and p-values for Task 2. In Part A (where participants used the prototypes), the pictorial-style prototype P2 slightly outperformed the textual list representations, improving scores by 0.53 points over P1 and 0.23 points over P3. Similarly, P4 improved scores by 0.36 points compared to P1 and 0.06 points compared to P3. When comparing P3 and P4, P4 showed a slight improvement of 0.06 points, though not statistically significant. Overall, the pictorial-style prototypes P2 and P4 led to higher scores than the textual list representations.

\begin{table}[ht]
\centering
\caption{Task 2. Comparison Results for PART A and PART B}
\renewcommand{\arraystretch}{1.2} 
\setlength{\tabcolsep}{4pt} 
\begin{tabularx}{\textwidth}{@{}lccccc|ccccc@{}}
\toprule
 & \multicolumn{5}{c}{\textbf{PART A}} & \multicolumn{5}{c}{\textbf{PART B}} \\ 
\cmidrule(lr){2-6} \cmidrule(lr){7-11}
\textbf{Pair} & \textbf{Coef} & \textbf{t} & \textbf{P>|t|} & \textbf{Corrected p} & \textbf{Cohen's d} & \textbf{Coef} & \textbf{t} & \textbf{P>|t|} & \textbf{Corrected p} & \textbf{Cohen's d} \\ 
\midrule
P1 vs P2 & 0.5330 & 1.520 & 0.133 & 0.798 & -0.353 & -0.0193 & -0.298 & 0.767 & 1 & 0.069 \\ 
P1 vs P3 & 0.3022 & 0.929 & 0.356 & 1 & -0.216 & -0.0872 & -1.463 & 0.148 & 0.888 & 0.340 \\ 
P1 vs P4 & 0.3631 & 1.071 & 0.288 & 1 & -0.254 & -0.0295 & -0.460 & 0.647 & 1 & 0.109 \\ 
P2 vs P3 & -0.2308 & -0.643 & 0.522 & 1 & 0.145 & -0.0679 & -1.132 & 0.261 & 1 & 0.256 \\ 
P2 vs P4 & -0.1699 & -0.455 & 0.651 & 1 & 0.105 & -0.0103 & -0.159 & 0.874 & 1 & 0.036 \\ 
P3 vs P4 & 0.0609 & 0.174 & 0.863 & 1 & -0.040 & 0.0577 & 0.970 & 0.335 & 1 & -0.224 \\ 
\bottomrule
\label{table:task2PairsStatisticalAnalysis}
\end{tabularx}
\end{table}

\begin{figure}[htbp]
    \centering
    \includegraphics[width=1\linewidth]{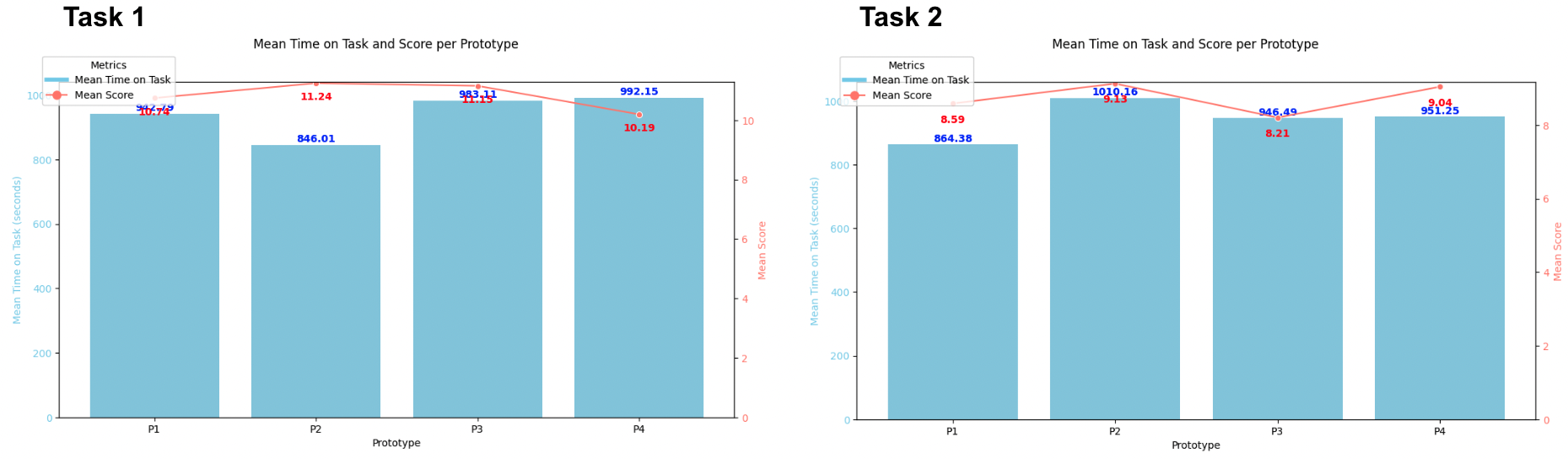}
    \caption{Teachers mean time on task and score per prototype}
    \label{fig:time_score}
\end{figure}

\textbf{Completion time:} Figure \ref{fig:time_score} summarises task completion times and mean scores per prototype. For Task 1, P2 reduced task completion time while maintaining score quality. In contrast, P3 and P4, which included contextual information (e.g., screenshots), required more time for exploration and interpretation. Notably, participants using P3 scored higher than those using P4, suggesting that excessive contextual information in P4 may have negatively impacted scores. For Task 2, P2 required more time, possibly due to the lack of contextual information, despite achieving the highest scores. P4, with screenshots, reduced task time compared to P2 while still improving scores over P3. This indicates that contextual elements in P4 facilitated task completion and improved performance, though P2 retained a slight edge in scores.

\begin{table}[h!]
\centering
\caption{Task completion time, comparison results}
\resizebox{\textwidth}{!}{%
\begin{tabular}{lcccccc|ccccc}
\toprule
\multirow{2}{*}{Pair} & \multicolumn{5}{c}{Task 1} & \multicolumn{5}{c}{Task 2} \\
\cmidrule(lr){2-6} \cmidrule(lr){7-11}
 & Coef & $t$ & P$>|t|$ & Corrected $p$ & Cohen’s $d$ & Coef & $t$ & P$>|t|$ & Corrected $p$ & Cohen’s $d$ \\
\midrule
P1 vs P2 & -65.0669 & -0.721 & 0.473 & 1 & 0.163 & 145.7845 & 1.383 & 0.171 & 1 & -0.321 \\
P1 vs P3 & 65.6553  & 0.563  & 0.575 & 1 & -0.127 & 82.1074  & 0.918  & 0.362 & 1 & -0.213 \\
P1 vs P4 & 66.7615  & 0.672  & 0.503 & 1 & -0.154 & 86.8702  & 1.025  & 0.309 & 1 & -0.243 \\
P2 vs P3 & 130.7222 & 1.180  & 0.241 & 1 & -0.263 & -63.6771 & -0.589 & 0.558 & 1 & 0.133 \\
P2 vs P4 & 131.8284 & 1.413  & 0.162 & 0.972 & -0.320 & -58.9143 & -0.553 & 0.582 & 1 & 0.127 \\
P3 vs P4 & 1.1062   & 0.009  & 0.993 & 1 & -0.002 & 4.7627   & 0.052  & 0.958 & 1 & 0.012 \\
\bottomrule
\end{tabular}%
}
\label{tab:CompletionTime_results}
\end{table}

For completion time, the comparison of coefficients, t-values, and p-values across prototypes for both tasks is summarised in Table \ref{tab:CompletionTime_results}. These results corroborate the findings depicted in Figure \ref{fig:time_score}, but none of the comparisons show statistical significance. 

\begin{figure}[htbp]
    \centering
    \includegraphics[width=.8\linewidth]{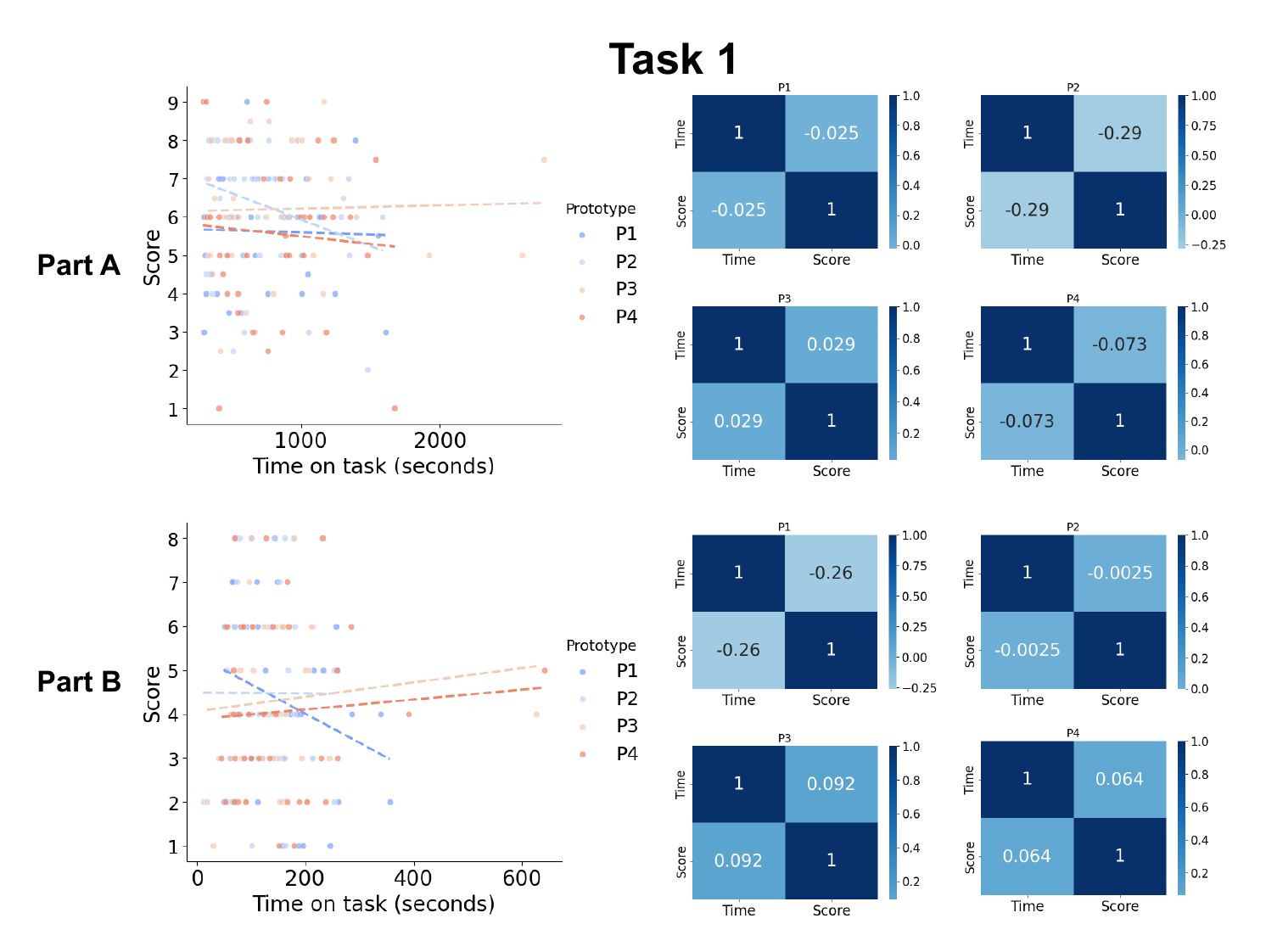}
    \caption{Task 1. Left: Time on Task vs Score by Prototypes. Right: Correlation heatmaps per prototype (parts A and B). 
    }
    \label{fig:Task1CorrTimeScore}
\end{figure}

\begin{figure}
    \centering
    \includegraphics[width=.8\linewidth]{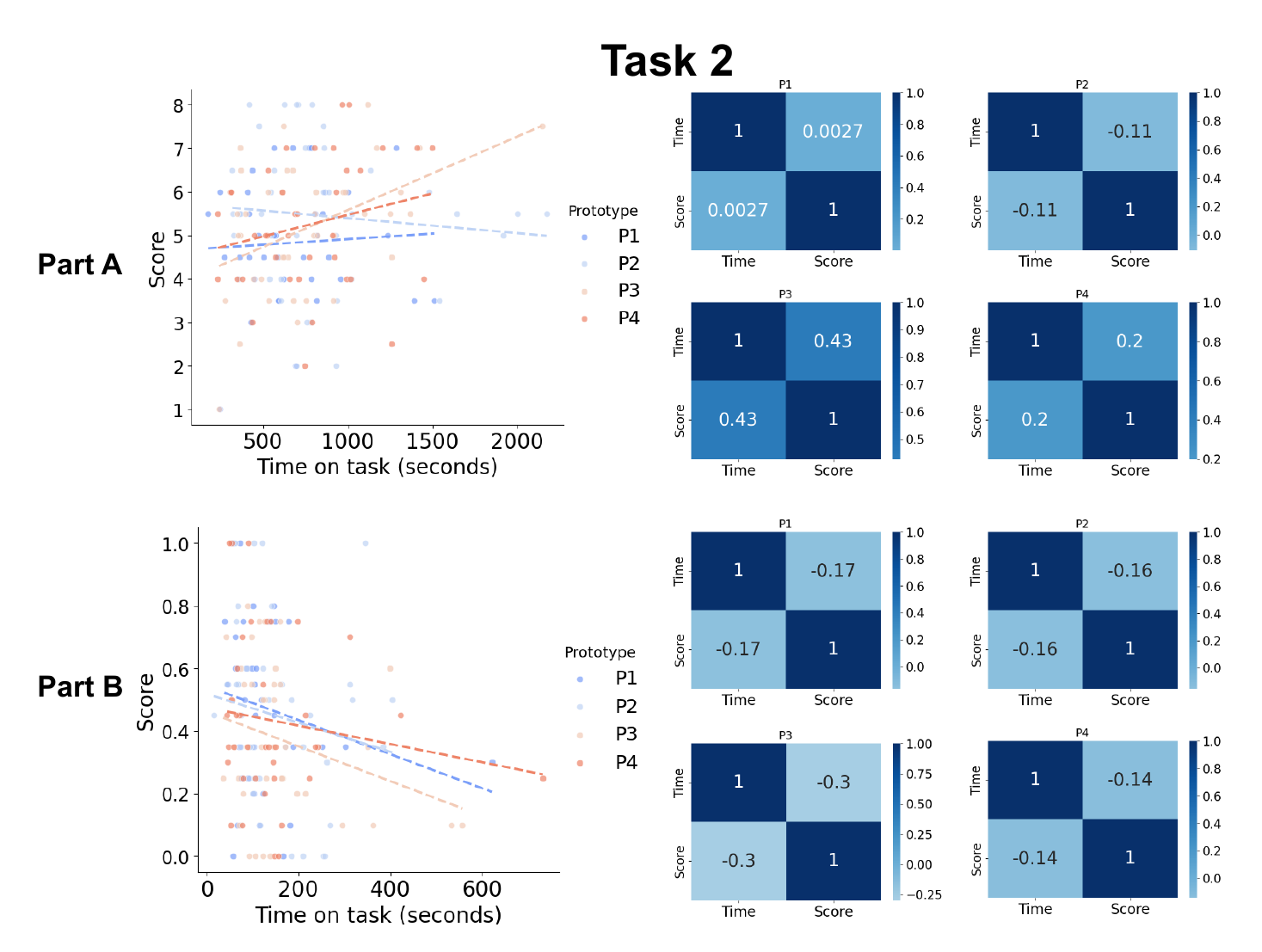}
    \caption{Task 2. Left: Time on Task vs Score by Prototypes. Right: Correlation heatmaps per prototype (parts A and B)}
    \label{fig:Task2CorrTimeScore}
\end{figure}

Figures \ref{fig:Task1CorrTimeScore} and \ref{fig:Task2CorrTimeScore} illustrate the correlation between time on task and scores for the four prototypes. In Part A of Task 1, all prototypes exhibit a negative trend, indicating that scores decrease as time on task increases. Only P3 shows a more stable or slightly improved trend. In Part B of Task 1, P3 and P4 demonstrated a positive trend, with P2 remaining stable and P1 continuing to show a negative trend. In Task 2, Part A, positive trends were observed for P1, P3, and P4, while P2 displays a negative trend, suggesting that more time on task is associated with lower scores for this prototype. For Part B, all prototypes exhibited negative trends, with P2 showing greater stability. The heatmaps presented in Figures \ref{fig:Task1CorrTimeScore} and \ref{fig:Task2CorrTimeScore} display the correlation coefficients between time on task and scores
across prototypes for both Parts A and B. In general, the heat maps indicate negative correlations between scores and time on the task, suggesting that participants who spent more time on the task tended to achieve lower scores. 

\textbf{Memorability:} Considering the outcomes of participants in Part B (see Table \ref{table:pairwiseTask1} for Task 1 and Table \ref{table:task2PairsStatisticalAnalysis} for Task 2), where they were not using the prototype and needed to recall elements of the process to answer the questions, the results showed the following trends: although no statistical significance was observed, the analysis indicated that participants using P2 achieved better task scores compared to P1 and P4. However, P1 outperformed P3 and P4, while P3 scored higher than P4. The comparison between P3 and P4 suggests that scaffolding contextual information in the representations may improve scores. This may also explain why P2 was more effective than P3: as a pictorial-style representation, P2 provided enough elements to support memorability, while P3, with its inclusion of contextual data (e.g., screenshots), may have introduced excessive information that increased cognitive load.

For task 2, Table \ref{table:task2PairsStatisticalAnalysis} reveals a pattern similar to Task 1. Participants who used P2 outperformed P3 and P4 on this task. However, participants who used P1 had better scores compared to those who used P2, P3, and P4. Notably, when comparing prototypes that include screenshots, P4 demonstrated positive effects on memorability compared to P3. This indicates that the way contextual information is integrated into pictorial-style representations, such as in P4, can play a critical role in supporting memorability.


In our exploration of memorability insights, Figures \ref{fig:Task1CorrTimeScore} and \ref{fig:Task2CorrTimeScore} present participants' time and scores without using the prototypes (part B), showing their performance trends. For Task 1, a positive trend was observed for P3 and P4, where increased time spent on the task was associated with slight improvement in scores. P2 was more stable and appeared to be more effective, as participants demonstrated quicker recall compared to other prototypes. In contrast, for P1, the trend indicates that the more time the participants spent on the task, the lower their scores were, suggesting inefficiency in this prototype. In Task 2 (Figure\ref{fig:Task2CorrTimeScore}, bottom), the results showed a different pattern. Here, the increase in time spent on the task generally correlated with lower scores across prototypes. However, P2 stood out, as it supported participants in completing the task more quickly, potentially indicating better memorability. 

\textbf{RQ2.} \textit{What are the differences in teachers' perceptions of usability and engagement across the four Visual Process Representations?}

\begin{figure}[htbp]
    \centering
    \includegraphics[width=1\linewidth]{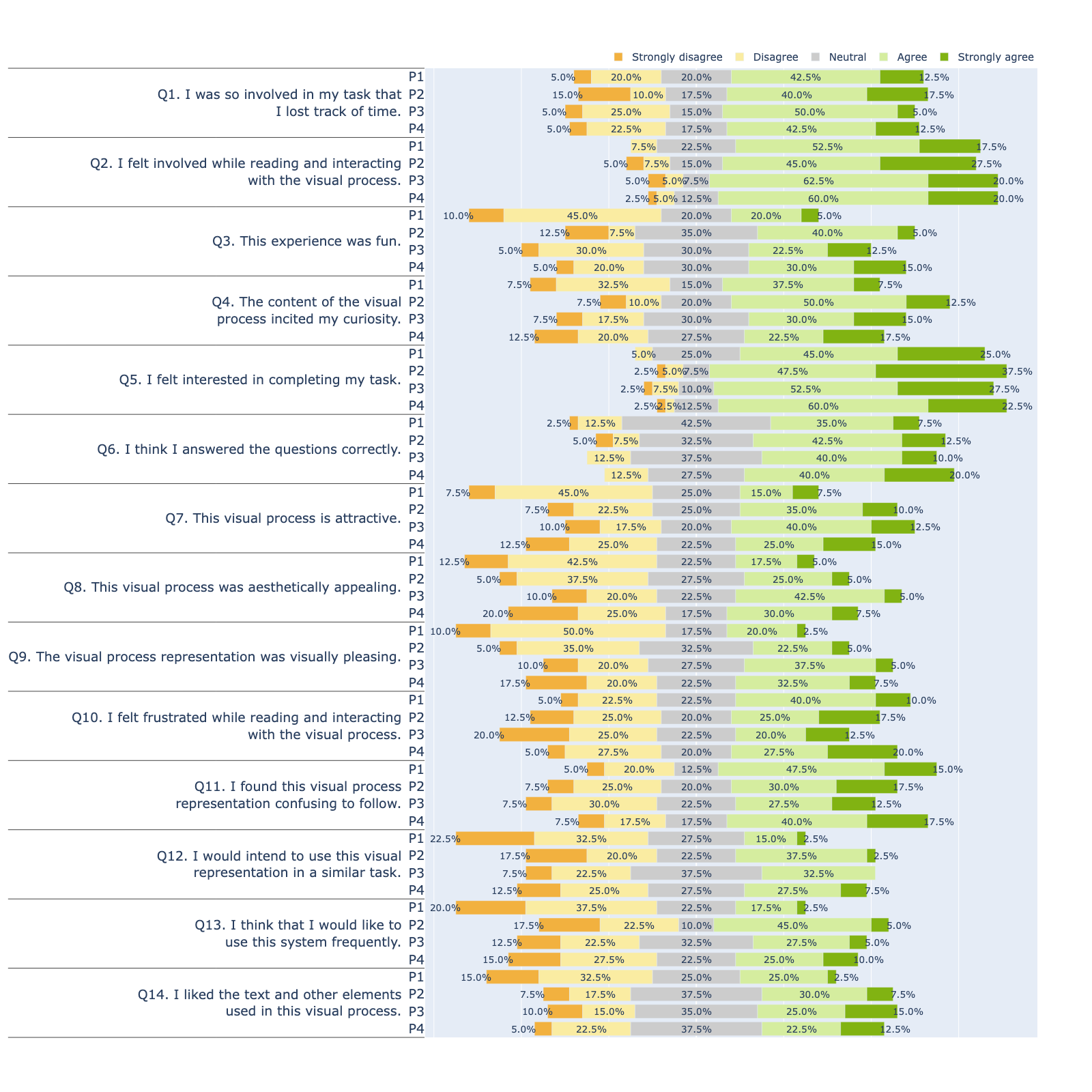}
    \caption{Teachers' engagement perceptions by prototype (P1-P4)}
    \label{fig:engagement}
\end{figure}

\begin{figure}[htbp]
    \centering
    \includegraphics[width=1\linewidth]{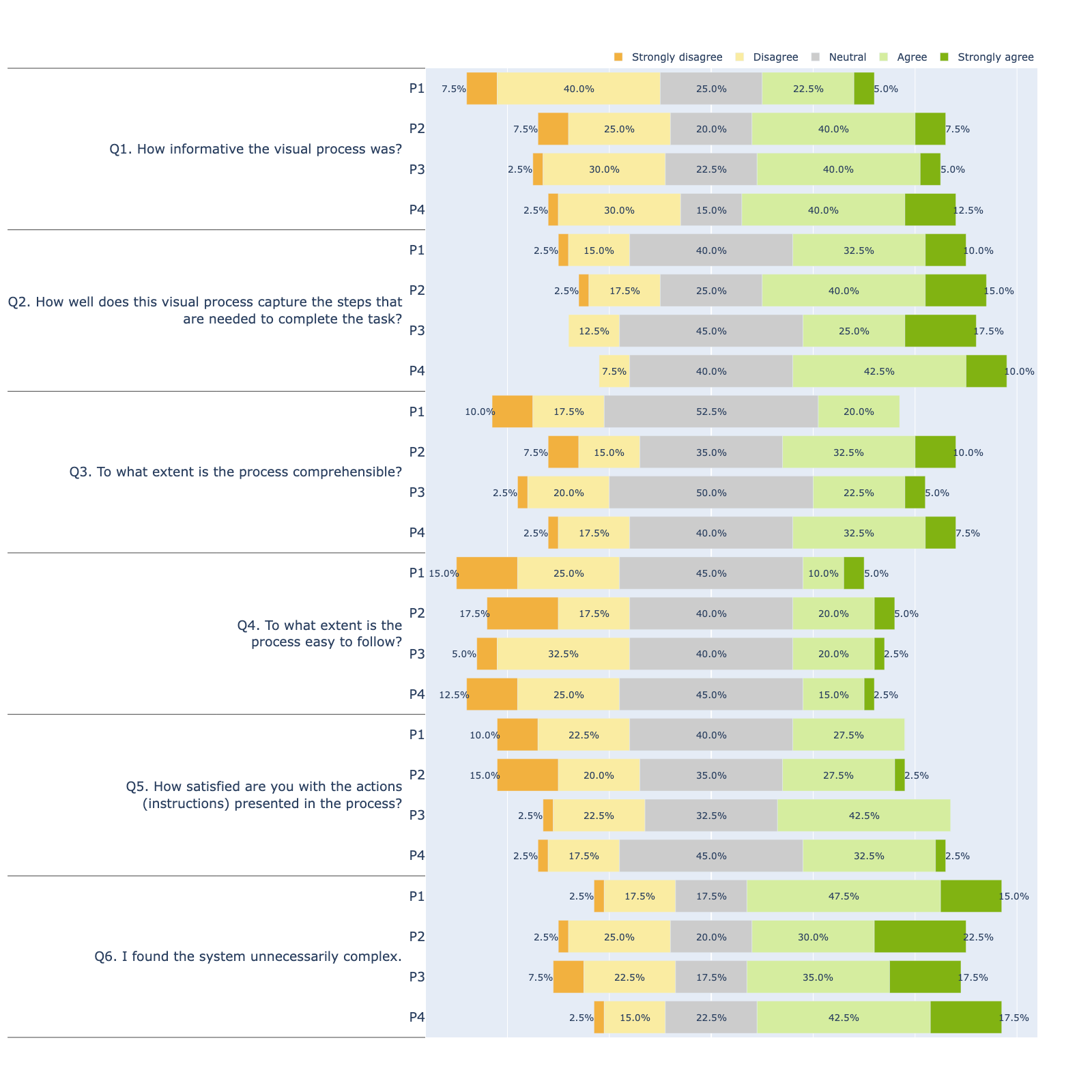}
    \caption{Teachers' usability perceptions by prototype (P1-P4)}
    \label{fig:sus}
\end{figure}

The engagement results (Figure \ref{fig:engagement}) highlight P2 and P4 as the most effective designs, with P4 slightly outperforming P2. P2 received high ratings for clarity, attractiveness, and task engagement, with participants agreeing or strongly agreeing that it was engaging (Q1: 57.5\%, Q5: 85\%, Q12: 40\%, Q13: 50\%), fun (Q3: 45.5\%), and curiosity-inducing (Q4: 62.5\%). P4 also excelled, particularly in task involvement (Q2: 80\%, Q5: 82.5\%, Q6: 60\%) and participant satisfaction (Q9: 40\%), though some expressed mild frustration (Q10: 47.5\%), suggesting areas for refinement. P3 received moderately positive ratings, scoring well on visual appeal (Q8: 47.5\%) and attractiveness (Q7: 52.5\%), but lacked the usability and polish of P2 and P4. In contrast, P1 scored lowest across usability metrics (Q1-Q9), with mostly neutral or negative responses, likely due to the absence of contextual information and engaging visual elements.

Although regression analysis was conducted for all engagement and usability questions across prototypes, only significant results are reported here. P4 received the highest ratings for fun and engagement (Q3, Figure \ref{fig:engagement}) with a moderate positive effect compared to P1 (t=2.648, p=0.009). P2 also showed significant improvement over P1 for the same question (t=2.138, p=0.034). For curiosity (Q4), P2, P3, and P4 were more engaging than P1.

Regarding attractiveness (Q7, Q8), P3 was rated the most visually appealing (coef=0.57, p=0.030), followed by P2 (coef=0.48, p=0.072). P3 also scored highest for aesthetic appeal (p=0.039). For reuse intentions (Q12), P3 (p=0.033) and P4 (p=0.043) were preferred over P1, with P2 showing promise (p=0.048). Participants favored the graphics in P3 (p=0.034) and P4 (p=0.055), although P3's simpler design was preferred. P4’s screenshots were appreciated but perceived as more complex. Overall, participants preferred visual elements in P3 and P4 over P1’s textual list of based steps.

In usability (Figure \ref{fig:sus}), P2 and P4 were rated the most effective, with P4 excelling in informativeness (Q1: 52.5\%), completeness (Q2: 52.5\%), comprehensibility (Q3: 42.5\%), and ease of following (Q4: 25\%). P2 performed well, capturing task steps effectively (Q2: 55\%), but received slightly more neutral responses. P3 received moderate feedback, with some neutrality on its ability to capture and convey information (Q3: 27.5\%, Q4: 22.5\%). P1 performed poorly, with high disagreement and neutral responses, indicating struggles with clarity and information delivery.

While P2 was positively received across many usability and engagement metrics, participants expressed that textual list of steps, as seen in P3, promoted less confusion during interaction. This suggests that the unfamiliar pictorial-style processes in P2 and P4 may cause initial challenges for first-time users.

\textbf{RQ3.} \textit{What improvements do teachers recommend for each Visual Process Representation prototype, and how do these suggestions align with identified engagement and usability challenges?} 

\textit{Opportunities for improvement:} In response to the question about the problems participants faced when navigating the different Visual Process Representations, we identified several opportunities to improve Visual Process Representations. Table \ref{tab:issues_summary} summarises the nine codes defined from the thematic analysis of responses from 160 participants. Specific codes highlight actionable areas for improving the prototypes. We will present selected quotes from the most prominent codes that illustrate teachers' (T1-T160) suggestions and insights for refinement. 

\begin{table}[h!]
\centering
\caption{Summary of codes defined based on participant responses across Visual Process Representations}
\label{tab:issues_summary}
\resizebox{\textwidth}{!}{%
\begin{tabular}{|l|c|c|c|c|c|}
\hline
\textbf{Code} & \textbf{P1} & \textbf{P2} & \textbf{P3} & \textbf{P4} & \textbf{Total responses} \\ \hline
Anxiety About Completing Task/Survey & 1 & 0 & 0 & 0 & 1 \\ \hline
Difficulty Distinguishing Between Steps/Subprocesses & 5 & 2 & 1 & 1 & 9 \\ \hline
Issues with the Study Design & 7 & 6 & 8 & 8 & 29 \\ \hline
Mixed Up Things Due to Working Too Fast & 1 & 0 & 0 & 0 & 1 \\ \hline
No Challenges Encountered & 12 & 17 & 16 & 21 & 66 \\ \hline
Prefer Simpler Representation/Written Manual & 1 & 0 & 0 & 1 & 2 \\ \hline
Suggestions for Improvement (Colours, Images, Simplification) & 8 & 5 & 5 & 2 & 20 \\ \hline
Terminology Used is Confusing/Non-Intuitive & 1 & 2 & 0 & 0 & 3 \\ \hline
Visual Representation Process is Complex/Confusing/Difficult/Overwhelming to Follow & 4 & 8 & 10 & 7 & 29 \\ \hline
\end{tabular}%
}
\end{table}

\textbf{Difficulty Distinguishing Between Steps/Subprocesses}: Referring to P1, T3 mentioned, \textit{"Due to the lack of subcategorisation, the elements under subprocesses seem repetitive. For example, if we want to shift to the next subprocess, which is distinct from the previous one, I’d recommend adding a distinctive title"}. Similarly, referring to P2, T65 stated, \textit{"I found it difficult to understand the difference between things the user would find and things the user should do"}. For P4, T154 noted, \textit{"The screenshots are too close together, making it slightly difficult to determine which step is next"}. Regarding P3, T118 mentioned, \textit{"Finding some of the next steps was challenging as they were not clearly labeled, like the Navigation"}. Although these comments were primarily about P1 (5 responses), they highlight a broader need for clearer identification of sections (subprocesses) within the Visual Process Representations and better guidance on navigating each section.

\textbf{Suggestions for Improvement (Colours, Images, Simplification)}: Regarding P1, T21 suggested, \textit{"The need for other colours (blue/purple). It would have been better to have a green navigation box and then change the colour for the added steps so it is clear that you have accomplished something—'yes, you have navigated, now do this'”}. Similarly, T27 noted, \textit{"Adding certain images between numerical action steps would be valuable for some people. Especially in the activity about the students' corrections, I would love to see part of the page as a still image. That would help me navigate faster and confirm the correctness of the information I have been given”}. This comment highlights the need for additional contextual data (e.g., screenshots). While the navigation was designed to use colour changes to indicate the current step, these comments suggest improvements to the colour scheme for subprocess navigation, providing better guidance on what has been completed and what needs to be done next.

Referring to P2, T54 mentioned, \textit{"I have to express that if one is not so accustomed to the order, one can get lost. I understand it follows the logic of a pictorial-style, but not everyone has that experience"}. This indicates the potential need for a familiarisation process for new users or those less accustomed to the pictorial-style format. T58 added, \textit{"I feel like a little bit more detail could be resourceful”}, further emphasising the need for contextual data (e.g., screenshots).

For P3, T93 noted, \textit{"The prototype was overly detailed, and it can be simplified"}, suggesting an opportunity to balance contextual information or provide it progressively. Finally, referring to P4, T158 recommended, \textit{"Consider adding more detailed descriptions or labels to each subprocess block, incorporating colour-coding for different types of actions, or including arrows to indicate the flow of the process"}.

\textbf{Terminology Used is Confusing/Non-Intuitive}: Although primarily related to P1 and P2, teachers' responses provide valuable insights for improving the terminology used in the Visual Process Representations. For P1, T10 noted, \textit{"The visualisation was useful, although more details were needed, and it was still difficult to process and relate the terminology used to the listed instructions"}. Referring to P2, T42 mentioned, \textit{"Understanding the terminology, initially, was a bit confusing. One of the explanatory diagrams included a label I was not sure I fully understood"}. These comments align with the code \textit{Difficulty Distinguishing Between Steps/Subprocesses}, highlighting the importance of using a common and intuitive language for steps and subprocesses that can be easily understood by both novices and experts.

\textbf{Visual Representation Process is Complex/Confusing/Difficult/Overwhelming to Follow}: For P1, T26 highlighted the need for \textit{"being focused during the task"} due to the prototype's navigation \textit{"complexity"} (T33) in completing tasks effectively. Additionally, T41 noted that the complexity might stem from the fact that they \textit{"personally need to do the actions at the same time to understand them"}. Regarding P2, and aligned with the previous comment, T47 mentioned that the complexity of the prototype was related to \textit{"having access to the actual interfaces, which would have been more intuitive to follow while using the visualisation"}. T46 added that P2 \textit{"was just overwhelming because the VPP was quite busy and detailed. It still needs more work to improve the UX"}. Other participants (T50, T56, T74) also described P2 navigation as complex. For P3, some teachers expressed visual discomfort, with T82 describing the prototype as \textit{"blurry to the eyes"}, and T100 noting that \textit{"although useful for completing the task, I felt unsure about how to follow it"}. In P4, T155 stated that \textit{"my cognitive load was quite heavy because I was trying to figure out the terminology for the different components of the Visual Process Representation, the names of the components, and what the components themselves represented"}. Similarly, T162 found the representation \textit{"really confusing overall. The way the processes were represented (both visually and in terms of the number of steps) seemed needlessly complex and unintuitive"}.

Twenty-nine out of 160 teachers provided responses associated with this code. While this represents a smaller subset of participants, it underscores the need for further iterations to address the complexity of the visualisations. For instance, it highlight the need for additional design iterations to simplify the representations, ensure common and intuitive terminology, and achieve a proper balance of steps, subprocesses, and contextual data (e.g., scaffolding). Improvements to layout and navigation are also necessary to enhance readability and ease of use across all prototypes. Finally, the fact that some teachers indicating the need for a real scenario to test the prototypes also suggested the need for an evaluation with a different type of task.  

\textbf{Other themes: } One participant reported feeling anxious while completing the task, but did not elaborate, preventing any conclusions. Twenty-nine teachers mentioned study design issues, primarily related to device limitations (e.g., screens too small to view all prototype information), which especially affected participants using contextual-data-rich prototypes (P3 and P4). Although the study was restricted to laptop users (not mobile devices), some technical aspects were beyond our control. Another common issue was participants’ unfamiliarity with the prototypes; although we provided an explanation and familiarisation time beforehand, offering additional hands-on interaction within a task context may help mitigate this in future studies.

\section{Discussion}

\subsection{Summary of the results}

For \textbf{RQ1}, results show that all prototypes supported teachers in completing tasks effectively, with P2 (pictorial-style without contextual information) achieving the highest task accuracy across both tasks, followed by P3. Although P4 (pictorial-style with contextual data) outperformed the textual prototype P1, the excessive contextual information in P4 increased cognitive load, negatively affecting performance and recall. Completion time analysis indicated that P2 enabled faster task completion while maintaining accuracy, particularly in Part B where memorability was crucial. These findings highlight the potential of pictorial-style representations—especially simpler versions like P2—for effectively communicating workflows. Our results align with prior work showing the benefits of visual storytelling techniques for enhancing workflow interpretability \cite{Dospan2023}. Similar approaches in other fields, such as medicine \cite{ComicMedicine2024} and energy/economy \cite{WangComicInfographic2019}, have demonstrated that pictorial-style visualisations can improve factual recall and comprehension compared to text or infographics. Although our study did not observe significant memorability improvements, these external findings suggest that with further refinement, pictorial elements hold promise for supporting knowledge transfer in educational settings. This study also demonstrates that automated methods can successfully transform log data into interpretable workflows by incorporating storytelling principles and inspiration from comic-narrative structures to support real teaching practices.

For \textbf{RQ2}, results reveal significant differences in teachers’ perceptions across prototypes. P2 consistently received the highest ratings for clarity, attractiveness, and engagement, demonstrating its ability to deliver a more intuitive and usable Visual Process Representation. P3 closely followed, excelling in visual appeal and participants’ willingness to reuse it, though some noted mild frustration due to complexity. P1 scored the lowest in usability and engagement metrics, reflecting challenges in delivering an engaging experience. Overall, pictorial-style representations (P2 and P4) enhanced engagement and usability, with P2 excelling in simplicity. Consistent with prior studies, pictorial-style visualisations have increased engagement in medical scenarios \cite{ComicMedicine2024} and public data communication \cite{WangComicInfographic2019}, where pictorial-style outperformed text representations and infographics in fostering engagement and comprehension.

For \textbf{RQ3}, results highlight key recommendations including simplifying navigation, improving clarity through colour coding and descriptive labels, and balancing contextual data to reduce cognitive load. Teachers also emphasised the importance of incorporating real-world scenarios (e.g., hands-on activities) in future evaluations. While our implementation followed prior visualisation recommendations \cite{WangComicInfographic2019}, such as structured layouts and navigation aids, challenges persist. For instance, as noted in our findings and by Wang et al. \cite{WangComicInfographic2019}, balancing contextual information, repetition, and sequencing is crucial, as excessive repetition or overloading information can confuse teachers. These findings underscore the need for further design refinements, such as highlighting patterns to improve navigation, employing innovative methods to scaffold contextual information, addressing usability challenges, and enabling immersive evaluations for KM tasks.

\subsection{Implications for Knowledge Management in Education}

The design approach and evaluation presented in this paper contribute significantly to KM in education by addressing the critical challenge of knowledge capture and transfer during staff turnover \cite{shah2020factors, azaki2022organisational}. As outlined in Section \ref{sec:designApproach}, our approach enables seamless knowledge capture from expert educators during their daily tasks. A key contribution is the automated generation of Visual Process Representations, which transform log data into interpretable artefacts that are easily understood, stored, shared, and reused by novice teachers. This aligns with Levallet and Chan \cite{levallet2019organizational}, who highlights that formalising knowledge in this manner facilitates knowledge sharing. Additionally, our findings support research showing that effective knowledge capture are critical for bridging knowledge sharing and use \cite{Demir02012023}. By integrating storytelling principles and drawing inspiration from comic-narrative structures, the Visual Process Representations enhance sharing, reuse, and long-term usability in educational contexts. Furthermore, this approach fosters an environment where educators are more inclined to share knowledge with colleagues and students, as suggested by prior KM research \cite{syed2021impact}.

Our approach reduces the time and effort required for manual documentation, allowing expert and novice teachers to focus more on essential activities such as improving learning design, which becomes increasingly important with rising university enrolments \cite{ujir2020teaching}. By alleviating the workload associated with documenting procedural tasks, the approach also addresses concerns raised by Meghan Stacey and McGrath-Champ \cite{Stacey02102022} about the growing administrative burden on educators. Visual Process Representations offer additional value by supporting infrequent yet critical tasks, such as semesterly administrative duties, ensuring that procedural knowledge remains accessible without diverting teachers from core teaching responsibilities.

This study highlights the pivotal role of technology in advancing KM strategies in education. While previous research has recognised the potential of digital tools for organisational knowledge management, our findings demonstrate how technology can seamlessly capture, codify, and transfer expert knowledge into accessible formats for novices.

\subsection{Implications for Visual Process Representations}

The evaluation results provide valuable insights into the visual design of Visual Process Representations, offering practical implications for improving usability, engagement, and effectiveness. Two main design considerations for Visual Process Representations emerged:

\textit{Leverage sequences in narrative visualisation:} Incorporating storytelling elements into visual representations, such as sequential panels and clear transitions, significantly enhances user engagement and comprehension of complex processes. Narrative techniques, including the temporal-sequence pattern defined by Bach et al. \cite{bachCHI2018Comicdesign}, depict temporal changes while guiding the reader's attention through a logical flow with a clear beginning, middle, and end. This structured storytelling approach ensures even complex processes are accessible and intuitive, aligning with findings from  Hullman et al. \cite{HullmanSequences}. By embedding these principles into Visual Process Representations, our designs offer a more engaging and cognitively manageable way to present workflows, reinforcing the importance of narrative in visualisation design.

\textit{Promote navigability and adaptability:} Balancing the inclusion of contextual information and visual complexity is essential to reducing cognitive load and enhancing usability. While simplified prototypes were better received overall, adding contextual data was beneficial for task completion but introduced challenges related to information overload. This reflects the findings of Stokes et al. \cite{StokesBalance2023} and Tewel et al. \cite{TewelTextVisual2023}, which emphasise user individual preferences such as preferring heavily annotated charts over simplified charts. To address these challenges, our qualitative results suggest adopting scaffolding strategies from data storytelling research, such as navigational alternatives used in comics \cite{InteractiveComicsWang2022} and colour-coded sections \cite{Dykes2015, Knaflic2017}, to provide intuitive guidance through complex workflows. Progressive sequencing (e.g., presenting one step or story at a time, as described by \cite{Martinez-Maldonado2020}) further enhances clarity and navigation. Aligning scaffolds with user needs ensures contextual information is delivered effectively without creating distractions.


These design considerations underscore the importance of iterative, user-centred development in refining visualisations. Insights from this study extend beyond educational contexts, informing the broader HCI and InfoVis communities on how users navigate and interpret narrative visualisations.

\subsection{Limitations and Future Work}

This study has several limitations that offer opportunities for future research. First, conducting the evaluation study on Prolific constrained the capture of nuanced user interaction data. While we collected click counts, we were unable to capture metrics such as scrolling behaviour, hover interactions, time spent on specific sections, or eye-tracking activity, which could provide deeper insights into user confusion or engagement and measure cognitive load. Future studies could integrate more advanced tracking mechanisms to better understand user behaviour.

Second, the prototypes were evaluated in controlled settings rather than real-world scenarios. This limits the ecological validity of the findings, as participants did not interact with the prototypes while completing authentic KM tasks. Future research should involve evaluations embedded within actual teaching workflows to assess the practical utility and scalability of Visual Process Representations. Future evaluations can also include understanding perceptions from the expert point of view and alternatives to capture teaching activities in class (physical or blended teaching tasks).

Third, while our approach demonstrated promise in improving task accuracy and usability, further iterations are needed to refine the prototypes. This includes optimising the balance between simplicity and contextual detail, enhancing memorability, and incorporating user feedback into subsequent designs. Additionally, future research could explore longitudinal studies to assess how Visual Process Representations impact knowledge retention and usability over time.

Finally, expanding the scope of application to other organisational contexts or educational institutions could validate the generalisability of the proposed approach. Exploring emerging technologies (e.g., interactive visualisations) or data collection in physical spaces (e.g., indoor positioning trackers), could further enhance the effectiveness of Visual Process Representations in supporting KM strategies.


\section{Conclusion}

In conclusion, this study highlights the transformative potential of combining automated methods with narrative visualisation to address critical Knowledge Management challenges in education. By seamlessly capturing teachers' workflows and translating them into intuitive Visual Process Representations, this approach ensures that tacit knowledge is captured and made accessible to novice educators. The findings demonstrate that well-designed visualisations, supported by thoughtful contextualisation and user-centred design principles, can significantly enhance task completion and engagement. Although limitations related to balancing contextual information, modelling more complex teaching tasks, and the use of controlled evaluation settings suggest areas for further research, this study provides a strong foundation for exploring scalable, flexible, and effective knowledge management solutions. Ultimately, this work underscores the vital role of visualisation in strengthening organisational memory and supporting institutions facing high teacher turnover.

\bibliographystyle{unsrt}  


\bibliography{bibliography}

\end{document}